\documentclass[epj]{svjour}
\usepackage[utf8]{inputenc}
\usepackage{amsmath,amssymb,amsbsy,bm,bbold,color,cite,hyperref}

\begin{document}
\title{Thermodynamics of a continuous medium with electric and magnetic dipoles}
\subtitle{}
\author{Sylvain D. Brechet\thanks{sylvain.brechet@epfl.ch} \and Jean-Philippe Ansermet\thanks{jean-philippe.ansermet@epfl.ch}
}                     
\institute{Institute of Condensed Matter Physics, Station 3, Ecole Polytechnique Fédérale de Lausanne - EPFL, CH-1015 Lausanne, Switzerland}
\date{Received: date / Revised version: date}

\abstract{
The thermodynamics of an electrically charged, multicomponent fluid with spontaneous electric and magnetic dipoles is analysed in the presence of electromagnetic fields. Taking into account the chemical composition of the current densities and stress tensors leads to three types of irreversible terms: scalars, vectors and pseudo-vectors. The scalar terms account for chemical reactivities, the vectorial terms account for transport and the pseudo-vectorial terms account for relaxation. The linear phenomenological relations, derived from the irreversible evolution, describe notably the Lehmann and electric Lehmann effects, the Debye relaxation of polar molecules and the Landau-Lifshitz relaxation of the magnetisation. This formalism accounts for the thermal and electric magnetisation accumulations and magnetisation waves. It also predicts that a temperature gradient affects the dynamics of magnetic vortices and drives magnetisation waves.
} 
\titlerunning{Thermodynamics of a continuous medium with electric and magnetic dipoles}
\authorrunning{Brechet et al.}

\maketitle
%


\section{Introduction}
\label{1}

Spin caloritronics is mainly focused on studying the effects of a temperature gradient on the time evolution of the distribution of the local spin average of a physical system~\cite{Bauer:2012}. In many experimental situations, the system can be treated as a classical continuum with magnetisation on the scale of interest where the quantum fluctuations average out and the underlying microscopic structure is smoothed out. In such as case, the local system is sufficiently large from a microscopic perspective to be treated classically, but it is sufficiently small from a macroscopic perspective to be considered as infinitesimal. In order to understand the interplay between the magnetisation and a temperature gradient, a consistent classical thermodynamic theory of continuous media with magnetisation needs to be established. It is of interest to include also the electric counterpart of the magnetisation, that is the electric polarisation. The thermodynamics of continuous media is presented in details in textbooks by Gurtin et al.~\cite{Gurtin:2010}, Wilmanski~\cite{Wilmanski:1998} and Silhavy~\cite{Silhavy:1997}. In order to account for the intrinsic rotation of the matter, we need to introduce explicitly an intrinsic angular velocity. The introduction of intrinsic rotations into the thermodynamics of continuous media was discussed notably by Müller~\cite{Mueller:1985} and Muschik~\cite{Muschik:2008b,Muschik:2001}. The electrodynamics of an electrically charged continuous medium interacting with electromagtnetic fields is detailled in textbooks by Eringen~\cite{Eringen:1990} and O'Dell~\cite{Odell:1970}. Some pioneering work on the non-equilibrium thermodynamics of a continuous medium with electromagnetic fields was carried out by Liu~\cite{Liu:1972}, Hutter~\cite{Hutter:1977} and de Groot and Mazur~\cite{deGroot:1984}. In the present work, we follow essentially the approach of Stückelberg detailed in reference~\cite{Stueckelberg:1974} and extend his formalism to describe locally the thermodynamics of an electrically charged continuous medium with spontaneous electric and magnetic dipoles in the presence of electromagnetic fields. 

In order to be able to define intensive fields such as the temperature, the chemical potential, the electrostatic potential, the electric field and the magnetic induction field, we require the local infinitesimal system to be homogeneous, uniform and at equilibrium.

We introduce phenomenological relations that express, in terms of the chemical constituents of the continuous medium, the mass density, the electric charge density, the intrinsic angular mass density, the electric polarisation and magnetisation, the momentum and the intrinsic angular momentum. Using these phenomenological relations, we express the electric polarisation and magnetisation current tensors as well as the stress and angular stress tensors in terms of the chemical current densities. We show that by expressing the state fields, such as the electric polarisation and the magnetisation, and the dynamical fields, such as the momentum density and the intrinsic angular momentum density, in terms of the different elementary constituents of the medium, three types of irreversible terms appear. First, there are scalar terms that account for the chemical reactivities and are expressed as the product between the chemical reaction rate densities and the chemical affinities. Second, there are vectorial terms that account for the transport and are expressed as the dot product between the current densities and the forces. Third, there are pseudo-vectorial terms that account for the relaxation and are expressed as the dot product between the intrinsic rotation rate densities and the intrinsic torques.

To illustrate our formalism, we derive explicit expressions for the irreversible relations characterising the Lehmann and electric Lehmann effects, and the relaxation of the electric and magnetic dipoles. Furthermore, we show how this formalism accounts for the thermal and electric magnetisation accumulations and magnetisation waves. We also predict how a temperature gradient affects the dynamics of magnetic vortices and drives magnetisation waves.

The structure of this publication is the following. In Sec.~\ref{Thermodynamic description}, we define the continuity equations, which describe the local time evolution for the material state variables. Then, we express the time evolution equations in terms of the different chemical constituents. In Sec.~\ref{Thermostatics and Thermodynamics}, using the fact that the energy is a state function, we combine the time evolution equations for the internal energy, the entropy and the other extensive material state variables. This ensures the local compatibility of the first and second laws of thermodynamics with the laws of mechanics. Through symmetry arguments, we establish the thermostatic, the reversible and irreversible thermodynamic equations. Finally, in Sec.~\ref{Thermodynamical phenomenology}, we establish the irreversible thermodynamic phenomenology and describe some physical phenomena to illustrate our formalism, with a particular emphasis on spintronics. 

\section{Thermodynamic description}
\label{Thermodynamic description}

\subsection{Continuity equations}

The state of a continuous medium is defined by a set of matter state variables. The local state of a continuous medium is defined by a set of matter state fields that are function of the space and time coordinates. To keep the notation concise, we do not explicitly write the space and time dependence of the state fields. The local thermodynamic state of a classical electrically charged continuous medium consisting of $N$ chemical substances with electric polarisation and magnetisation interacting with electromagnetic fields is determined by the following state fields:
\begin{itemize}
\item[$\bullet$] the entropy density field $s$,
\item[$\bullet$] the number density fields $n_A$ of $N$ chemical substances, where $A\in\{1,\ldots,N\}$,
\item[$\bullet$] the electric charge density field $q$,
\item[$\bullet$] the spontaneous electric polarisation field $\mathbf{P}$,
\item[$\bullet$] the spontaneous magnetisation field $\mathbf{M}$,
\item[$\bullet$] the velocity field $\mathbf{v}$,
\item[$\bullet$] the intrinsic angular velocity field $\bm{\omega}$.
\end{itemize}
It is important to stress that the system does not include the electromagnetic fields in the vacuum, but only their interaction with the matter fields. Thus, the state fields $q$, $\mathbf{P}$ and $\mathbf{M}$, accounting for the electromagnetic properties of the matter, are purely matter fields, in contrast to our previous work~\cite{Brechet:2012,Brechet:2013} where the electric displacement field $\mathbf{D}$ and the magnetic induction field $\mathbf{B}$ were accounting for the electromagnetic properties of the matter and of the electromagnetic fields in the vacuum. 

Moreover, in this publication, we restrict our analysis to a system with spontaneous electric polarisation $\mathbf{P}$ and magnetisation $\mathbf{M}$, whereas in our former work~\cite{Brechet:2012,Brechet:2013}, we considered only a system with induced electric polarisation and magnetisation.

The classical time evolution of every extensive physical property in a local infinitesimal system is expressed by a continuity equation, which is a local detailed balance equation. In this publication, we shall refer to the continuity equations expressed in terms of the material time derivative as the material continuity equations.

The material continuity equation for an extensive scalar property $F$ is given by~\cite{Brechet:2012},
\begin{equation}\label{cont eq f sca}
\dot{f} + \left(\bm{\nabla}\cdot\mathbf{v}\right)f + \bm{\nabla}\cdot\mathbf{j}_f = \rho_f\ ,
\end{equation}
where $\dot{f}$ is the material time derivative of the scalar density $f$, $\mathbf{j}_f$ is the vectorial diffusive current density and $\rho_f$ is the scalar source density of $F$. Note that the frame-independent material time derivative operator is related to the partial time derivative operator by,
\begin{equation*}
\dot{\vphantom{\rho}}\, = \partial_t + \left(\mathbf{v}\cdot\bm{\nabla}\right)\ .
\end{equation*}

The material continuity equation for an extensive vectorial property $\mathbf{F}$ is given by~\cite{Brechet:2012},
\begin{equation}\label{cont eq f vec}
\bm{\dot{\mathbf{f}}} + \left(\bm{\nabla}\cdot\mathbf{v}\right)\mathbf{f} + \bm{\nabla}\cdot\mathsf{j}_{\mathbf{f}} = \bm{\rho}_{\mathbf{f}}\ ,
\end{equation}
where $\bm{\dot{\mathbf{f}}}$ is the material time derivative of the vectorial density $\mathbf{f}$, $\mathsf{j}_{\mathbf{f}}$ is the rank-$2$ tensorial diffusive current density and $\bm{\rho}_{\mathbf{f}}$ is the vectorial source density of $\mathbf{F}$.

The material continuity equation~\eqref{cont eq f sca} for the entropy yields,
\begin{equation}\label{cont eq S}
\dot{s} + \left(\bm{\nabla}\cdot\mathbf{v}\right)s + \bm{\nabla}\cdot\mathbf{j}_s = \rho_s\geqslant 0\ ,
\end{equation}
where $s$ is the entropy density, $\mathbf{j}_s$ is the diffusive entropy current density and $\rho_s$ is the entropy source density that is positive-definite in order to satisfy locally the second law of thermodynamics. Note that the strictly positive part, i.e. $\rho_s>0$, accounts for the irreversibility by breaking the symmetry of the dynamics under time inversion, which defines a time arrow. 

The material continuity equation~\eqref{cont eq f sca} for the number of elementary units of chemical substance $A$ is given by,
\begin{equation}\label{cont eq A}
\dot{n}_A + \left(\bm{\nabla}\cdot\mathbf{v}\right)n_A + \bm{\nabla}\cdot\mathbf{j}_A = \sum_a\,\nu_{aA}\,\omega_a\ ,
\end{equation}
where $n_A$ is the number density of the chemical substance $A$, $\mathbf{j}_A$ is the diffusive current density and the source density $\rho_A$ is expressed explicitly as the sum over all the chemical reactions $a$ of the product of the stoichiometric coefficients $\nu_{aA}$ and the reaction rate densities $\omega_a$.

The material continuity equation~\eqref{cont eq f sca} for the electric charge is given by,
\begin{equation}\label{cont eq q}
\dot{q} + \left(\bm{\nabla}\cdot\mathbf{v}\right)q + \bm{\nabla}\cdot\mathbf{j}_q = 0\ ,
\end{equation}
where $q$ is the electric charge density, $\mathbf{j}_q$ is the diffusive electric current density and there is no electric charge source density $\rho_q$ since the electric charge is an invariant. Note that the electric current density $\mathbf{j}$ is the sum of the convective and diffusive electric current densities, i.e.
\begin{equation}\label{current j}
\mathbf{j} = q\,\mathbf{v} + \mathbf{j}_q\ .
\end{equation}

The material continuity equation~\eqref{cont eq f vec} for the spontaneous electric dipoles is given by,
\begin{equation}\label{cont eq pol}
\bm{\dot{\mathbf{P}}} + \left(\bm{\nabla}\cdot\mathbf{v}\right)\mathbf{P} + \bm{\nabla}\cdot\mathsf{j}_{\mathbf{P}} = \sum_A\,\bm{\Omega}_A^{\,\text{m}}\times\mathbf{p}_A\ ,
\end{equation}
where $\mathbf{P}$ is the spontaneous electric polarisation, $\mathsf{j}_{\mathbf{P}}$ is the diffusive electric polarisation current tensor. The source density $\rho_{\mathbf{P}}$ of the spontaneous electric dipoles $\mathbf{p}_A$ of the chemical substances $A$ is an irreversible term accounting for the relaxation of the spontaneous electric dipoles $\mathbf{p}_A$. It is due to the rotational motion of the chemical substances $A$, that is expressed in terms of the intrinsic rotation rate densities $\bm{\Omega}_A^{\,\text{m}}$ of the matter where the superscript $^{\,\text{m}}$ stands for ``matter''. 

The uniformity of the local infinitesimal system implies that there is a unique local rotation rate density $\bm{\Omega}_A^{\,\text{m}}$ for each chemical substance $A$, that is defined with respect to the local frame where the electric polarisation $\mathbf{P}$ does not rotate. This frame corresponds to the local frame where the intrinsic rotation of the matter vanishes, i.e. $\bm{\omega}=\mathbf{0}$, since the spontaneous electric polarisation is rigidly attached to the matter.
 
The material continuity equation~\eqref{cont eq f vec} for the spontaneous magnetic dipoles yields,
\begin{equation}\label{cont eq mag}
\begin{split}
&\bm{\dot{\mathbf{M}}} + \left(\bm{\nabla}\cdot\mathbf{v}\right)\mathbf{M} + \bm{\nabla}\cdot\mathsf{j}_{\mathbf{M}} =\\
&\sum_A\,\Big(\gamma_A\,n_A\left(\mathbf{m}_A\times\mathbf{B}\right) + \bm{\Omega}_A^{\,\mathbf{M}}\times\mathbf{m}_A\Big)\ ,
\end{split}
\end{equation}
where $\mathbf{M}$ is the spontaneous magnetisation, $\mathsf{j}_{\mathbf{M}}$ is the diffusive magnetisation current tensor. The source density $\rho_{\mathbf{M}}$ of the spontaneous magnetic dipoles $\mathbf{m}_A$ of the chemical substances $A$ consists of two terms. The first term is a reversible term accounting for the precession of the magnetic dipoles $\mathbf{m}_A$ around the magnetic induction field $\mathbf{B}$ that is expressed in terms of their gyromagnetic ratios $\gamma_A$. The second term is an irreversible term accounting for the relaxation of the magnetic dipoles $\mathbf{m}_A$ with respect to the magnetic induction field $\mathbf{B}$ that is expressed in terms of the intrinsic rotation rate densities $\bm{\Omega}_A^{\,\mathbf{M}}$ where the superscript $^{\,\mathbf{M}}$ stands for ``magnetisation''. These rotation rate densities consist of two contributions. The first contribution corresponds to the micromagnetic hypothesis~\cite{Brown:1962}, that accounts for the fact that electrons are point-like particles that have no known substructure and thus no intrinsic angular mass. The second contribution accounts for the rotational motion of the chemical substances $A$. The second contribution is generally much smaller than the first. 

The uniformity of the local infinitesimal system implies that there is a unique local rotation rate density $\bm{\Omega}_A^{\,\mathbf{M}}$ for each chemical substance $A$, that is defined with respect to the local frame where the magnetisation $\mathbf{M}$ does not rotate. In contrast to the case of the electric polarisation, this frame does not correspond to the local frame where the intrinsic rotation of the matter vanishes, since the spontaneous magnetisation is not rigidly attached to the matter. Note that by symmetry, only axial vectors such as the magnetic dipoles $\mathbf{m}_A$ are allowed to precess. Thus, no precession of the electric dipoles $\mathbf{p}_A$ is to be expected. 

The material continuity equation~\eqref{cont eq f vec} for the momentum is given by,
\begin{equation}\label{cont eq mom}
\bm{\dot{\mathbf{p}}} + \left(\bm{\nabla}\cdot\mathbf{v}\right)\mathbf{p} -\,\bm{\nabla}\cdot\bm{\sigma} = \mathbf{f}^{\,\text{ext}}\ ,
\end{equation}
where $\mathbf{p}$ is the momentum density, $\bm{\sigma}$ is the stress tensor and $\mathbf{f}^{\,\text{ext}}$ is the external force density. The momentum density $\mathbf{p}\left(n_A,\mathbf{v}\right)$ of the matter is proportional to the velocity $\mathbf{v}$, i.e.
\begin{equation}\label{momentum 0}
\mathbf{p}\left(n_A,\mathbf{v}\right) = m\left(n_A\right)\,\mathbf{v}\ ,
\end{equation}
where $m\left(n_A\right)$ is the mass density.

The material continuity equation~\eqref{cont eq f sca} for the mass yields,
\begin{equation}\label{cont eq mass}
\dot{m} + \left(\bm{\nabla}\cdot\mathbf{v}\right)m = 0\ .
\end{equation}
There is no diffusive mass current density $\mathbf{j}_m$ by definition of the local centre of mass and there is no mass source density $\rho_m$ since the mass is a non-relativistic invariant. Substituting the relation~\eqref{momentum 0} and the continuity equation~\eqref{cont eq mass} for the mass into the continuity equation~\eqref{cont eq mom} for the momentum, the latter yields Newton's second law of motion, i.e.
\begin{equation}\label{Newton law}
m\,\bm{\dot{\mathbf{v}}} = \mathbf{f}^{\,\text{ext}} + \bm{\nabla}\cdot\bm{\sigma}\ .
\end{equation}

The material continuity equation~\eqref{cont eq f vec} for the intrinsic angular momentum yields,
\begin{equation}\label{cont eq ang mom}
\bm{\dot{\mathbf{s}}} + \left(\bm{\nabla}\cdot\mathbf{v}\right)\mathbf{s} -\,\bm{\nabla}\cdot\bm{\Theta} = \bm{\tau}^{\,\text{ext}}\ ,
\end{equation}
where $\mathbf{s}$ is the intrinsic angular momentum density, $\bm{\Theta}$ is the intrinsic angular stress tensor and $\bm{\tau}^{\,\text{ext}}$ is the intrinsic torque density. The local infinitesimal system is considered as a homogeneous sphere. Thus, the intrinsic angular momentum density $\mathbf{s}\left(n_A,\bm{\omega}\right)$ of the matter is proportional to the intrinsic angular velocity $\bm{\omega}$, i.e.
\begin{equation}\label{angular momentum 0}
\mathbf{s}\left(n_A,\bm{\omega}\right) = I\left(n_A\right)\,\bm{\omega}\ ,
\end{equation}
where $I\left(n_A\right)$ is the intrinsic angular mass density, which is a scalar density. For a homogeneous sphere of constant infinitesimal radius $dr$, the intrinsic angular mass density $I$ is related to the mass density $m$ by,
\begin{equation}\label{int mass mass}
I = \frac{2}{5}\,m\,dr^2\ .
\end{equation}

Thus, the material continuity equation~\eqref{cont eq mass} for the mass and the relation~\eqref{int mass mass} imply that the material continuity equation~\eqref{cont eq f sca} for the intrinsic angular mass is given by,
\begin{equation}\label{cont eq intrinsic angular mass}
\dot{I} + \left(\bm{\nabla}\cdot\mathbf{v}\right)I = 0\ .
\end{equation}
There is no diffusive intrinsic angular mass current density $\mathbf{j}_I$ and no intrinsic angular mass source density $\rho_I$ since for a homogeneous local system the intrinsic angular mass density $I$ is totally determined by the mass density $m$.

Substituting the relation~\eqref{angular momentum 0} and the continuity equation~\eqref{cont eq intrinsic angular mass} for the intrinsic angular mass into the continuity equation~\eqref{cont eq ang mom} for the intrinsic angular momentum, the latter yields Newton's second law in intrinsic rotation, i.e.
\begin{equation}\label{Euler eq}
I\,\bm{\dot{\omega}} = \bm{\tau}^{\,\text{ext}} + \bm{\nabla}\cdot\bm{\Theta}\ .
\end{equation}

The material continuity equation~\eqref{cont eq f sca} for the energy is given by,
\begin{equation}\label{cont eq energy}
\dot{e} + \left(\bm{\nabla}\cdot\mathbf{v}\right)e + \bm{\nabla}\cdot\mathbf{j}_e = \mathbf{f}^{\,\text{ext}}\cdot\mathbf{v} + \bm{\tau}^{\,\text{ext}}\cdot\bm{\omega}\ ,
\end{equation}
where $e$ is the energy density, $\mathbf{j}_e$ is the diffusive energy current density and the energy source density $\rho_{e}$ is the power density due to the external force densities $\mathbf{f}^{\,\text{ext}}$ and external intrinsic torque densities $\bm{\tau}^{\,\text{ext}}$. The energy density $e\left(s,n_A,q,\mathbf{P},\mathbf{M},\mathbf{v},\bm{\omega}\right)$ is the sum of the translational kinetic energy density, i.e. $\frac{1}{2}\,m\,\mathbf{v}^2$, the intrinsic rotational kinetic energy density, i.e. $\frac{1}{2}\,I\,\bm{\omega}^2$, and the internal energy density, i.e. $u\left(s,n_A,q,\mathbf{P},\mathbf{M}\right)$,
\begin{equation}\label{energy 0}
\begin{split}
&e\left(s,n_A,q,\mathbf{P},\mathbf{M},\mathbf{v},\bm{\omega}\right) = \frac{1}{2}\,m\,\mathbf{v}^2 + \frac{1}{2}\,I\,\bm{\omega}^2\\
&\phantom{e\left(s,n_A,q,\mathbf{P},\mathbf{M},\mathbf{v},\bm{\omega}\right) = } + u\left(s,n_A,q,\mathbf{P},\mathbf{M}\right)\ ,
\end{split}
\end{equation}
where the internal energy density is the energy density in the local rest frame where $\mathbf{v} = \mathbf{0}$ and $\bm{\omega} = \mathbf{0}$. Using the continuity equations for the mass~\eqref{cont eq mass} and the intrinsic angular mass~\eqref{cont eq intrinsic angular mass}, Newton's second law in translation~\eqref{Newton law} and in intrinsic rotation~\eqref{Euler eq}, the time derivative of the relation~\eqref{energy 0} yields,
\begin{align}\label{energy deriv}
&\dot{e} = \dot{u} + \left(\bm{\nabla}\cdot\mathbf{v}\right)\left(u -\,e\right) + \left(\bm{\nabla}\cdot\bm{\sigma}\right)\cdot\mathbf{v} + \left(\bm{\nabla}\cdot\bm{\Theta}\right)\cdot\bm{\omega}\nonumber\\
&\phantom{\dot{e} = }+ \mathbf{f}^{\,\text{ext}}\cdot\mathbf{v} + \bm{\tau}^{\,\text{ext}}\cdot\bm{\omega}\ .
\end{align}
Substituting the relation~\eqref{energy deriv} into the continuity equation~\eqref{cont eq energy} for the energy and using the vectorial identities 
\begin{align*}
&\left(\bm{\nabla}\cdot\bm{\sigma}\right)\cdot\mathbf{v} = \bm{\nabla}_{\bm{\sigma}}\cdot\left(\bm{\sigma}\cdot\mathbf{v}\right) - \bm{\sigma}\cdot\left(\bm{\nabla}\odot\mathbf{v}\right)\ ,\\
&\left(\bm{\nabla}\cdot\bm{\Theta}\right)\cdot\bm{\omega} = \bm{\nabla}_{\bm{\Theta}}\cdot\left(\bm{\Theta}\cdot\bm{\omega}\right) - \bm{\Theta}\cdot\left(\bm{\nabla}\odot\bm{\omega}\right)\ ,
\end{align*}
yields the material continuity equation~\eqref{cont eq f sca} for the internal energy density, i.e.
\begin{equation}\label{cont eq int energy}
\dot{u} + \left(\bm{\nabla}\cdot\mathbf{v}\right)u + \bm{\nabla}\cdot\mathbf{j}_u = \bm{\sigma}\cdot\left(\bm{\nabla}\odot\mathbf{v}\right) + \bm{\Theta}\cdot\left(\bm{\nabla}\odot\bm{\omega}\right)\ ,
\end{equation}
where the symbol $\odot$ denotes a symmetrised tensorial product $\otimes$ and the indices $\vphantom{\bm{\nabla}}_{\bm{\sigma}}$ or $\vphantom{\bm{\nabla}}_{\bm{\Theta}}$ denote that there is a dot product between the covariant differential operator $\bm{\nabla}$ and the contravariant components of the stress tensors $\bm{\sigma}$ and $\bm{\Theta}$ respectively.

The internal energy current density $\mathbf{j}_u$ is related to the energy current density $\mathbf{j}_e$ by,
\begin{equation}\label{j_u}
\mathbf{j}_e = \mathbf{j}_u -\,\bm{\sigma}\cdot\mathbf{v} -\,\bm{\Theta}\cdot\bm{\omega}\ .
\end{equation}
Note that the source density terms $\bm{\sigma}\cdot\left(\bm{\nabla}\odot\mathbf{v}\right)$ and $\bm{\Theta}\cdot\left(\bm{\nabla}\odot\bm{\omega}\right)$ in the continuity equation~\eqref{cont eq int energy} for the internal energy account for the elastic expansion or contraction of the matter and for the dissipative translational and rotational viscous friction. The dissipation is defined as the irreversible part of the internal energy source density $\rho_u$.

\subsection{Dynamics in terms of the chemical composition}

\subsubsection{Time evolution of the electric charge density}
The electric charge density $q$ is defined as the product of the number density $n_A$ and the electric charge $q_A$ carried by the elementary units of chemical substances $A$~\cite{Brechet:2012}, i.e.
\begin{equation}\label{electric charge density}
q = \sum_A\,n_A\,q_A\ .
\end{equation}
In order to characterise physically the time evolution of the electric polarisation, we substitute the expression~\eqref{electric charge density} for the chemical composition of the electric charge density and the continuity equation for the chemical substance~\eqref{cont eq A} into the continuity equation~\eqref{cont eq q} for the electric charge. The latter then becomes,
\begin{equation}\label{cont eq charge bis}
\sum_A\Big(-\,\left(\bm{\nabla}\cdot\mathbf{j}_A\right)\,q_A + \sum_a\,\nu_{aA}\,\omega_a\,q_A\Big) + \bm{\nabla}\cdot\mathbf{j}_q = 0\ .
\end{equation}
Using the fact that the electric charge $q_A$ is an invariant, i.e.
\begin{equation*}
\left(\bm{\nabla}\cdot\mathbf{j}_A\right)\,q_A = \bm{\nabla}\cdot\left(q_A\,\mathbf{j}_A\right)\ ,
\end{equation*}
the continuity equation~\eqref{cont eq charge bis} for the electric charge can be recast as,
\begin{align}\label{cont eq charge ter}
\sum_{a,A}\,\nu_{aA}\,\omega_a\,q_A + \bm{\nabla}\cdot\Big(\mathbf{j}_q -\,\sum_A\,q_A\,\mathbf{j}_A\Big) = 0\ .
\end{align}
The continuity equation~\eqref{cont eq charge ter} has to hold for any current flow, which yields an explicit expression for the diffusive electric current density, i.e.
\begin{equation}\label{el charge cur dens}
\mathbf{j}_q = \sum_A\,q_A\,\mathbf{j}_A\ .
\end{equation}
Moreover, it has to hold for every chemical reaction $a$, which implies that
\begin{equation}\label{chem charge}
\sum_{A}\,\nu_{aA}\,q_A = 0\ ,
\end{equation}
and means that the chemical reaction $a$ preserves the total electric charge in the local infinitesimal system. 

\subsubsection{Time evolution of the electric dipoles}
The electric polarisation $\mathbf{P}$ is defined as the product of the number density $n_A$ and the electric dipoles $\mathbf{p}_A$ carried by the elementary units of chemical substances $A$, i.e.
\begin{equation}\label{electric polarisation}
\mathbf{P} = \sum_A\,n_A\,\mathbf{p}_A\ .
\end{equation}
In order to characterise physically the dynamics of the electric polarisation, we substitute the expression~\eqref{electric polarisation} for the chemical composition of the electric polarisation and the continuity equation for the chemical substance~\eqref{cont eq A} into the continuity equation~\eqref{cont eq pol} for the electric dipoles. The latter then becomes,
\begin{align}\label{cont eq pol bis}	
&\sum_A\Big(n_A\,\bm{\dot{\mathbf{p}}}_A -\,\left(\bm{\nabla}\cdot\mathbf{j}_A\right)\mathbf{p}_A + \sum_a\,\nu_{aA}\,\omega_a\,\mathbf{p}_A\Big) + \bm{\nabla}\cdot\mathsf{j}_{\mathbf{P}}\nonumber\\
&= \sum_A\,\bm{\Omega}_A^{\,\text{m}}\times\mathbf{p}_A\ .
\end{align}
Using the vectorial identity,
\begin{equation*}
\left(\bm{\nabla}\cdot\mathbf{j}_A\right)\mathbf{p}_A = \bm{\nabla}_{\mathbf{j}}\cdot\left(\mathbf{p}_A\odot\mathbf{j}_A\right) -\,\left(\mathbf{j}_A\cdot\bm{\nabla}\right)\mathbf{p}_A\ ,
\end{equation*}
where the index $\vphantom{\bm{\nabla}}_{\mathbf{j}}$ denotes that there is a dot product between the covariant differential operator $\bm{\nabla}$ and the contravariant current density $\mathbf{j}_A$,
the continuity equation~\eqref{cont eq pol bis} for the electric dipoles can be recast as,
\begin{equation}\label{cont eq pol ter}
\begin{split}	
&\sum_A\Big(n_A\,\bm{\dot{\mathbf{p}}}_A -\,\bm{\Omega}_A^{\,\text{m}}\times\mathbf{p}_A + \left(\mathbf{j}_A\cdot\bm{\nabla}\right)\mathbf{p}_A\Big)\\
&+ \sum_{a,A}\,\nu_{aA}\,\omega_a\,\mathbf{p}_A + \bm{\nabla}_{\mathbf{j}}\cdot\Big(\mathsf{j}_{\mathbf{P}} -\,\sum_A\,\mathbf{p}_A\odot\mathbf{j}_A\Big) = \mathbf{0}\ .
\end{split}
\end{equation}
The continuity equation~\eqref{cont eq pol ter} has to hold for any current flow, which yields an explicit expression for the electric polarisation current tensor, i.e.
\begin{equation}\label{el pol cur dens}
\mathsf{j}_{\mathbf{P}} = \sum_A\,\mathbf{p}_A\odot\mathbf{j}_A\ .
\end{equation}
The condition~\eqref{el pol cur dens} implies that the continuity equation~\eqref{cont eq pol ter} reduces to,
\begin{equation}\label{cont eq pol quad}
\sum_A\Big(n_A\,\bm{\dot{\mathbf{p}}}_A -\,\bm{\Omega}_A^{\,\text{m}}\times\mathbf{p}_A + \left(\mathbf{j}_A\cdot\bm{\nabla}\right)\mathbf{p}_A + \sum_a\,\nu_{aA}\,\omega_a\,\mathbf{p}_A\Big) = \mathbf{0}\ ,
\end{equation}
which describes the time evolution of the electric dipoles $\mathbf{p}_A$ of all the chemical substances.

The local time evolution of the electric dipoles $\mathbf{p}_A$ of a specific chemical substance $A$ is given by,
\begin{equation}\label{eq el pol}
n_A\,\bm{\dot{\mathbf{p}}}_A = -\,\mathbf{p}_A\times\bm{\Omega}_A^{\,\textbf{m}} -\,\left(\mathbf{j}_A\!\cdot\!\bm{\nabla}\right)\mathbf{p}_A -\,\sum_a\,\nu_{aA}\,\omega_a\,\mathbf{p}_A\ ,
\end{equation}
where the first, second and third terms on the RHS describe respectively the relaxation, the transport and the chemistry of the electric dipoles $\mathbf{p}_A$. Note the intrinsic rotation rate densities $\bm{\Omega}_A^{\,\textbf{m}}$ are functions of the electric dipoles of all the chemical substances. They account for the dissipative couplings between the electric dipoles of the different chemical substances.

\subsubsection{Time evolution of the magnetic dipoles}
The magnetisation $\mathbf{M}$ is defined as the product of the number density $n_A$ and the magnetic dipoles $\mathbf{m}_A$ carried by the elementary units of chemical substances $A$, i.e.
\begin{equation}\label{magnetisation}
\mathbf{M} = \sum_A\,n_A\,\mathbf{m}_A\ .
\end{equation}
In order to characterise physically the time evolution of the magnetisation, we substitute the expression~\eqref{magnetisation} for the chemical composition of the magnetisation and the continuity equation for the chemical substance~\eqref{cont eq A} into the continuity equation~\eqref{cont eq mag} for the magnetic dipoles. The latter then becomes,
\begin{align}\label{cont mag bis}	
&\sum_A\Big(n_A\,\bm{\dot{\mathbf{m}}}_A -\,\left(\bm{\nabla}\cdot\mathbf{j}_A\right)\mathbf{m}_A + \sum_a\,\nu_{aA}\,\omega_a\,\mathbf{m}_A\Big) + \bm{\nabla}\cdot\mathsf{j}_{\mathbf{M}}\nonumber\\
&= \sum_A\,\Big(\gamma_A\,n_A\,\mathbf{m}_A\times\mathbf{B} + \bm{\Omega}_A^{\,\mathbf{M}}\times\mathbf{m}_A\Big)\ .
\end{align}
Using the vectorial identity,
\begin{equation*}
\left(\bm{\nabla}\cdot\mathbf{j}_A\right)\mathbf{m}_A = \bm{\nabla}_{\mathbf{j}}\cdot\left(\mathbf{m}_A\odot\mathbf{j}_A\right) -\,\left(\mathbf{j}_A\cdot\bm{\nabla}\right)\mathbf{m}_A\ ,
\end{equation*}
the continuity equation~\eqref{cont mag bis} for the magnetic dipoles can be recast as,
\begin{equation}\label{cont mag ter}
\begin{split}
&\sum_A\Big(n_A\left(\bm{\dot{\mathbf{m}}}_A-\gamma_A\,\mathbf{m}_A\times\mathbf{B}\right) -\bm{\Omega}_A^{\,\mathbf{M}}\!\times\mathbf{m}_A + \left(\mathbf{j}_A\cdot\bm{\nabla}\right)\mathbf{m}_A\Big)\\
&+ \sum_{a,A}\,\nu_{aA}\,\omega_a\,\mathbf{m}_A + \bm{\nabla}_{\mathbf{j}}\cdot\Big(\mathsf{j}_{\mathbf{M}} -\,\sum_A\,\mathbf{m}_A\odot\mathbf{j}_A\Big) = \mathbf{0}\ .
\end{split}
\end{equation}
The continuity equation~\eqref{cont mag ter} has to hold for any current flow, which yields an explicit expression for the magnetisation current tensor, i.e.
\begin{equation}\label{mag cur dens}
\mathsf{j}_{\mathbf{M}} = \sum_A\,\mathbf{m}_A\odot\mathbf{j}_A\ .
\end{equation}
Note that this expression for the magnetisation current tensor is the classical counterpart of the tensorial ``spin-current'' used in spintronics~\cite{Stiles:2002}. The condition~\eqref{mag cur dens} implies that the continuity equation~\eqref{cont mag ter} reduces to,
\begin{equation}\label{cont mag quad}
\begin{split}
&\sum_A\Big(n_A\left(\bm{\dot{\mathbf{m}}}_A-\,\gamma_A\,\mathbf{m}_A\times\mathbf{B}\right) -\,\bm{\Omega}_A^{\,\mathbf{M}}\times\mathbf{m}_A\\
&\phantom{\sum_A\ \ }+ \left(\mathbf{j}_A\cdot\bm{\nabla}\right)\mathbf{m}_A + \sum_{a}\,\nu_{aA}\,\omega_a\,\mathbf{m}_A\Big) = \mathbf{0}\ ,
\end{split}
\end{equation}
which describes the local time evolution of the magnetic dipoles of all the chemical substances. 

In order to establish the local time evolution of the magnetic dipoles $\mathbf{m}_A$ of a specific chemical substance $A$ in the presence of other chemical substances $B$ carrying magnetic dipoles $\mathbf{m}_B$, we need to take explicitly into account the local magnetic torque that couples $\mathbf{m}_A$ and $\mathbf{m}_B$. This torque is antisymmetric under the permutation of different chemical substances $A$ and $B$, which implies that the torque density vanishes after summation over all the chemical substances, i.e.
\begin{equation}\label{int mag}
\sum_{A,B}\,\gamma_{AB}\,n_A\,n_B\left(\mathbf{m}_A\times\mathbf{m}_B\right) = \mathbf{0}\ ,
\end{equation}
where $\gamma_{AB}$ a the symmetric coupling coefficient. Note that the magnetic torques do not affects the magnetisation $\mathbf{M}$, which is defined as the density of magnetic dipoles of all the chemical substances. Thus, they do not affect the Larmor energy of the local system. Note that by symmetry, the interaction torques are allowed only for axial vectors such as the magnetic dipoles $\mathbf{m}_A$ and $\mathbf{m}_B$. Thus, no interaction torque of the electric dipoles $\mathbf{p}_A$ and $\mathbf{p}_B$ is to be expected. 

The equations~\eqref{cont mag quad} and~\eqref{int mag} imply that the local time evolution of the magnetic dipoles $\mathbf{m}_A$ of a specific chemical substance $A$ is given by,
\begin{align}\label{eq mag}
&n_A\,\bm{\dot{\mathbf{m}}}_A = \gamma_A\,n_A\,\left(\mathbf{m}_A\times\mathbf{B}\right) -\,\mathbf{m}_A\times\bm{\Omega}_A^{\,\mathbf{M}} -\,\left(\mathbf{j}_A\!\cdot\!\bm{\nabla}\right)\mathbf{m}_A\nonumber\\
&\phantom{n_A\,\bm{\dot{\mathbf{m}}}_A } -\,\sum_{a}\,\nu_{aA}\,\omega_a\,\mathbf{m}_A + \sum_{B}\,\gamma_{AB}\,n_A\left(\mathbf{m}_A\times n_B\,\mathbf{m}_B\right)\,,
\end{align}
where the first, second, third and fourth terms on the RHS describe respectively the precession, the relaxation, the transport and the chemistry of the magnetic dipoles $\mathbf{m}_A$, and the last term describes the local interaction of the magnetic dipoles $\mathbf{m}_A$ with the magnetic dipoles $\mathbf{m}_B$ of the other chemical substances $B$. Note the intrinsic rotation rate densities $\bm{\Omega}_A^{\,\mathbf{M}}$ are functions of the magnetic dipoles of all the chemical substances. They account for the dissipative couplings between the magnetic dipoles of the different chemical substances.

\subsubsection{Time evolution of the mass density}
The mass density $m$ is defined as the product of the number density $n_A$ and the mass $m_A$ carried by the elementary units of chemical substances $A$~\cite{Brechet:2012}, i.e.
\begin{equation}\label{mass density}
m = \sum_A\,n_A\,m_A\ .
\end{equation}
In order to characterise physically the time evolution of the mass density, we substitute the expression~\eqref{mass density} for the chemical composition of the mass density and the continuity equation for the chemical substance~\eqref{cont eq A} into the continuity equation~\eqref{cont eq mass} for the mass. The latter then becomes,
\begin{equation}\label{cont eq mass bis}
\sum_A\Big(-\,\left(\bm{\nabla}\cdot\mathbf{j}_A\right)\,m_A + \sum_a\,\nu_{aA}\,\omega_a\,m_A\Big) = 0\ .
\end{equation}
Using the fact that the mass $m_A$ is a non-relativistic invariant, i.e.
\begin{equation*}
\left(\bm{\nabla}\cdot\mathbf{j}_A\right)\,m_A = \bm{\nabla}\cdot\left(m_A\,\mathbf{j}_A\right)\ ,
\end{equation*}
the continuity equation~\eqref{cont eq mass bis} for the mass can be recast as,
\begin{align}\label{cont mass ter}
\sum_{a,A}\,\nu_{aA}\,\omega_a\,m_A -\,\bm{\nabla}\cdot\Big(\sum_A\,m_A\,\mathbf{j}_A\Big) = 0\ .
\end{align}
The continuity equation~\eqref{cont mass ter} has to hold for any current flow, which defines a set of frames where the local centre of mass of the matter element is at rest, i.e.
\begin{equation}\label{mass cur dens}
\sum_A\,m_A\,\mathbf{j}_A = \mathbf{0}\ .
\end{equation}
Moreover, it has to hold for every chemical reaction $a$, which implies that
\begin{equation}\label{chem mass}
\sum_{A}\,\nu_{aA}\,m_A = 0\ ,
\end{equation}
and means that the chemical reaction $a$ preserves the total mass in the local infinitesimal system. 

\subsubsection{Time evolution of the momentum}
The momentum density $\mathbf{p} = m\,\mathbf{v}$ is defined as the density of momentum carried by the elementary units of chemical substances $A$ of velocities $\mathbf{v}_A$, i.e.
\begin{equation}\label{momentum}
\mathbf{p} = \sum_A\,n_A\,m_A\,\mathbf{v}_A\ .
\end{equation}
It consists of convective part and a diffusive parts according to,
\begin{equation}\label{velocities}
\sum_A\,n_A\,m_A\,\mathbf{v}_A = \sum_A\,n_A\,m_A\,\mathbf{v} + \sum_A\,m_A\,\mathbf{j}_A\ .
\end{equation}
Thus, using the relation~\eqref{mass cur dens}, the velocity of the matter element is found to be,
\begin{equation}
\label{velocity}
\mathbf{v} = \left(\sum_A\,n_A\,m_A\right)^{-1}\!\left(\sum_A\,n_A\,m_A\,\mathbf{v}_A\right)\ ,
\end{equation}
and represents the velocity of the local centre of mass.

In order to characterise physically the time evolution of the momentum, we substitute the expression~\eqref{momentum} for the chemical composition of the momentum density and the continuity equation for the chemical substance~\eqref{cont eq A} into the continuity equation~\eqref{cont eq mom} for the momentum. The latter then becomes,
\begin{align}\label{cont mom bis}
&\sum_A\Big(n_A\,m_A\,\bm{\dot{\mathbf{v}}}_A -\,\left(\bm{\nabla}\cdot\mathbf{j}_A\right)m_A\,\mathbf{v}_A + \sum_a\,\nu_{aA}\,\omega_a\,m_A\,\mathbf{v}_A\Big)\nonumber\\
&-\,\bm{\nabla}\cdot\bm{\sigma} = \mathbf{f}^{\,\text{ext}}\ .
\end{align}
Using the vectorial identity,
\begin{equation*}
\left(\bm{\nabla}\cdot\mathbf{j}_A\right)\mathbf{v}_A = \bm{\nabla}_{\mathbf{j}}\cdot\left(\mathbf{v}_A\odot\mathbf{j}_A\right) -\,\left(\mathbf{j}_A\cdot\bm{\nabla}\right)\mathbf{v}_A\ ,
\end{equation*}
and the extensivity of the force, i.e.
\begin{equation*}
\mathbf{f}^{\,\text{ext}} = \sum_A\,\mathbf{f}^{\,\text{ext}}_{A}\ ,
\end{equation*}
where $\mathbf{f}^{\,\text{ext}}_{A}$ represents the external force density acting on the chemical substance $A$, the continuity equation~\eqref{cont mom bis} for the momentum can be recast as,
\begin{align}\label{cont mom ter}
&\!\sum_A\Big(n_A\,m_A\,\bm{\dot{\mathbf{v}}}_A + m_A\,\left(\mathbf{j}_A\cdot\bm{\nabla}\right)\mathbf{v}_A -\,\mathbf{f}^{\,\text{ext}}_{A}\Big)\\
&\!+ \sum_{a,A}\,\nu_{aA}\,\omega_a\,m_A\,\mathbf{v}_A -\,\bm{\nabla}_{\mathbf{j}}\cdot\Big(\bm{\sigma} + \sum_A\,m_A\,\mathbf{v}_A\odot\mathbf{j}_A\Big) = \mathbf{0}\,.\nonumber
\end{align}
The matter stress tensor $\bm{\sigma}$ is split into a reversible part due to the pressure $P$ and an irreversible part $\bm{\tilde{\sigma}}$ according to,
\begin{equation}\label{sigma}
\bm{\sigma} = -\,P\,\mathbb{1} + \bm{\tilde{\sigma}}\ ,
\end{equation}
which implies that the continuity equation~\eqref{cont mom ter} is recast as,
\begin{align}\label{cont mom ter 2}
&\!\!\sum_A\Big(n_A\,m_A\,\bm{\dot{\mathbf{v}}}_A + m_A\,\left(\mathbf{j}_A\cdot\bm{\nabla}\right)\mathbf{v}_A -\,\mathbf{f}^{\,\text{ext}}_{A}\Big) + \bm{\nabla}\,P \\
&\!\!+ \sum_{a,A}\,\nu_{aA}\,\omega_a\,m_A\,\mathbf{v}_A -\,\bm{\nabla}_{\mathbf{j}}\cdot\Big(\bm{\tilde{\sigma}} + \sum_A\,m_A\,\mathbf{v}_A\odot\mathbf{j}_A\Big) = \mathbf{0}\,.\nonumber
\end{align}

The continuity equation~\eqref{cont mom ter 2} has to hold for any current flow, which yields an explicit expression for the irreversible part of the stress tensor, i.e.
\begin{equation}\label{mom cur dens}
\bm{\tilde{\sigma}} = -\,\sum_A\,m_A\,\mathbf{v}_A\odot\mathbf{j}_A\ .
\end{equation}
Moreover, every chemical reaction preserves the total momentum in the local infinitesimal system, i.e.
\begin{equation}\label{chem mom}
\sum_A\,\nu_{aA}\,m_A\,\mathbf{v}_A = \mathbf{0}\ .
\end{equation}

Taking into account the fact that the pressure $P$ of the continuous medium is the sum of the partial pressures $P_A$ of the different chemical substances $A$ multiplied by the dimensionless fugacity coefficients $\varphi_A$~\cite{OConnell:2005}, i.e.
\begin{equation}\label{pressure}
P = \sum_A\,\varphi_A\,P_A\ ,
\end{equation}
and using the conditions~\eqref{mom cur dens} and~\eqref{chem mom}, the continuity equation~\eqref{cont mom ter 2} reduces to,
\begin{equation}\label{cont mom quad}
\sum_A\Big(n_A\,m_A\,\bm{\dot{\mathbf{v}}}_A + m_A\,\left(\mathbf{j}_A\cdot\bm{\nabla}\right)\mathbf{v}_A -\,\mathbf{f}^{\,\text{ext}}_{A} + \bm{\nabla}\left(\varphi_A\,P_A\right)\Big) = \mathbf{0}\,,
\end{equation}
which describes the local time evolution of all the chemical substances. Note that the dimensionless fugacity coefficient $\varphi_A$ in the relation~\eqref{pressure} accounts for the local interactions between the constituents of the chemical substance $A$.

In order to establish the local time evolution of a specific chemical substance $A$ in the presence of other chemical substances $B$, we need to take explicitly into account the local internal force densities $\mathbf{f}^{\,\text{int}}_{B\rightarrow A}$ exerted by the chemical substances $B$ on the chemical substance $A$. These force densities do not change the local internal energy of the system, which means that they are the densities of conservative forces accounting for elastic collisions or scattering. Newton's third law implies that the sum of the internal force densities over all the chemical substances vanishes, i.e.
\begin{equation}\label{int mom}
\sum_{A,B}\,\mathbf{f}^{\,\text{int}}_{B\rightarrow A} = \mathbf{0}\ .
\end{equation}
The equations~\eqref{cont mom quad} and~\eqref{int mom} imply that the local time evolution of a specific chemical substance $A$ is given by,
\begin{equation}\label{eq mom}
n_A\,m_A\,\bm{\dot{\mathbf{v}}}_A = \mathbf{f}^{\,\text{ext}}_{A} -\,\bm{\nabla}\left(\varphi_A\,P_A\right) -\,m_A\left(\mathbf{j}_A\!\cdot\!\bm{\nabla}\right)\mathbf{v}_A + \sum_{B}\,\mathbf{f}^{\,\text{int}}_{B\rightarrow A}\ ,
\end{equation}
where the first, second and third terms on the RHS describe respectively the action of the external forces, of the fugacity gradient and of the transport, and the last term describes the action of the internal forces due to the local interaction of the chemical substance $A$ with the other chemical substances $B$.

\subsubsection{Time evolution of the intrinsic angular momentum}
The intrinsic angular momentum density $\mathbf{s} = I\,\bm{\omega}$ is defined as the density of intrinsic angular momentum carried by the elementary units of chemical substances $A$ of intrinsic angular velocities $\bm{\omega}_A$, i.e.
\begin{equation}\label{angular momentum}
\mathbf{s} = \sum_A\,n_A\,I_A\,\bm{\omega}_A\ ,
\end{equation}
where $n_A\,I_A$ represent the intrinsic angular mass density of the elementary units of chemical substance $A$ and the relations~\eqref{int mass mass} and~\eqref{mass density} imply that
\begin{equation}\label{int ang mass bis}
n_A\,I_A = \frac{2}{5}\,n_A\,m_A\,dr^2\ .
\end{equation}
The intrinsic angular momentum density $\mathbf{s}$ consists of convective and diffusive parts according to,
\begin{equation}\label{angular velocities}
\sum_A\,n_A\,I_A\,\bm{\omega}_A = \sum_A\,n_A\,I_A\,\bm{\omega} + \sum_A\,I_A\,\bm{\Omega}_A^{\,\text{m}}\ ,
\end{equation}
where the sum of the diffusive intrinsic angular momenta of all the chemical substances $A$ vanish, i.e.
\begin{equation}\label{intrinsic angular mass rest cond}
\sum_A\,I_A\,\bm{\Omega}_A^{\,\text{m}} = \mathbf{0}\ ,
\end{equation}
which defines a set of frames where the matter has no average intrinsic rotational motion. Thus, the intrinsic angular velocity of the matter element is found to be,
\begin{equation}
\label{angular velocity}
\bm{\omega} = \left(\sum_A\,n_A\,I_A\right)^{-1}\left(\sum_A\,n_A\,I_A\,\bm{\omega}_A\right)\ ,
\end{equation}
and represents the average intrinsic angular velocity around the local centre of mass.

In order to characterise physically the time evolution of the intrinsic angular momentum, we substitute the expression~\eqref{angular momentum} for the chemical composition of the intrinsic angular momentum density and the continuity equation for the chemical substance~\eqref{cont eq A} into the continuity equation~\eqref{cont eq ang mom} for the intrinsic angular momentum. The latter then becomes,
\begin{align}\label{cont ang mom bis}
&\sum_A\Big(n_A\,I_A\,\bm{\dot{\omega}}_A -\,\left(\bm{\nabla}\cdot\mathbf{j}_A\right)I_A\,\bm{\omega}_A + \sum_a\,\nu_{aA}\,\omega_a\,I_A\,\bm{\omega}_A\Big)\nonumber\\
&-\,\bm{\nabla}\cdot\bm{\Theta} = \bm{\tau}^{\,\text{ext}}\ .
\end{align}
Using the vectorial identity,
\begin{equation*}
\left(\bm{\nabla}\cdot\mathbf{j}_A\right)\bm{\omega}_A = \bm{\nabla}_{\mathbf{j}}\cdot\left(\bm{\omega}_A\odot\mathbf{j}_A\right) -\,\left(\mathbf{j}_A\cdot\bm{\nabla}\right)\bm{\omega}_A\ ,
\end{equation*}
and the extensivity of the torque, i.e.
\begin{equation*}
\bm{\tau}^{\,\text{ext}} = \sum_A\,\bm{\tau}^{\,\text{ext}}_{A}\ ,
\end{equation*}
where $\bm{\tau}^{\,\text{ext}}_{A}$ represents the external torque density acting on the chemical substance $A$, the continuity equation~\eqref{cont ang mom bis} for the intrinsic angular momentum is recast as,
\begin{align}\label{cont ang mom ter}
&\sum_A\Big(n_A\,I_A\,\bm{\dot{\omega}}_A + I_A\left(\mathbf{j}_A\cdot\bm{\nabla}\right)\bm{\omega}_A -\,\bm{\tau}^{\,\text{ext}}_{A}\Big)\\
& + \sum_{a,A}\,\nu_{aA}\,\omega_a\,I_A\,\bm{\omega}_A -\,\bm{\nabla}_{\mathbf{j}}\cdot\Big(\bm{\Theta} + \sum_A\,I_A\,\bm{\omega}_A\odot\mathbf{j}_A\Big) = \mathbf{0}\ .\nonumber
\end{align}

The continuity equation~\eqref{cont ang mom ter} has to hold for any current flow, which yields an explicit expression for the irreversible part of the intrinsic angular stress tensor, i.e.
\begin{equation}\label{ang mom cur dens}
\bm{\Theta} = -\,\sum_A\,I_A\,\bm{\omega}_A\odot\mathbf{j}_A\ .
\end{equation}
Moreover, every chemical reaction preserves the total intrinsic angular momentum in the local infinitesimal system, i.e.
\begin{equation}\label{chem ang mom}
\sum_A\,\nu_{aA}\,I_A\,\bm{\omega}_A = \mathbf{0}\ .
\end{equation}

The conditions~\eqref{ang mom cur dens} and~\eqref{chem ang mom} imply that the continuity equation~\eqref{cont ang mom ter} reduces to,
\begin{equation}\label{cont ang mom quad}
\sum_A\Big(n_A\,I_A\,\bm{\dot{\omega}}_A + I_A\left(\mathbf{j}_A\cdot\bm{\nabla}\right)\bm{\omega}_A -\,\bm{\tau}^{\,\text{ext}}_{A}\Big) = \mathbf{0}\ ,
\end{equation}
which describes the local intrinsic time evolution of all the chemical substances. 

In order to establish the local intrinsic time evolution of a specific chemical substance $A$ in the presence of other chemical substances $B$, we need to take explicitly into account the local internal intrinsic torque densities $\bm{\tau}^{\,\text{int}}_{B\rightarrow A}$ exerted by the chemical substances $B$ on the chemical substance $A$. These torque densities do not change the local internal energy of the system, which means that they are the densities of conservative torques accounting for elastic collisions or scattering. Newton's third law in intrinsic rotation implies that the sum of the internal torque densities over all the chemical substances vanishes, i.e.
\begin{equation}\label{int ang mom}
\sum_{A,B}\,\bm{\tau}^{\,\text{int}}_{B\rightarrow A} = \mathbf{0}\ .
\end{equation}
The equations~\eqref{cont ang mom quad} and~\eqref{int ang mom} imply that the local intrinsic time evolution of a specific chemical substance $A$ is given by,
\begin{equation}\label{eq ang mom}
n_A\,I_A\,\bm{\dot{\bm{\omega}}}_A = \bm{\tau}^{\,\text{ext}}_{A} -\,I_A\left(\mathbf{j}_A\cdot\bm{\nabla}\right)\bm{\omega}_A + \sum_{B}\,\bm{\tau}^{\,\text{int}}_{B\rightarrow A}\ ,
\end{equation}
where the first and second terms on the RHS describe respectively the action of the external torques and of the transport, and the last term describes the action of the internal torques due to the local interaction of the chemical substance $A$ with the other chemical substances $B$.

\section{Thermostatics and Thermodynamics}
\label{Thermostatics and Thermodynamics}
\subsection{Thermostatic equation, reversible and irreversible thermodynamic equations}

The thermostatics and thermodynamics of the continuous medium are contained within the internal energy balance~\eqref{cont eq int energy}. Since the internal energy density $u\left(s,n_A,q,\mathbf{P},\mathbf{M}\right)$ is a state function in the local rest frame, the time derivative of the internal energy density field is given by, i.e.
\begin{equation}\label{deriv u}
\dot{u} = T\,\dot{s} + \sum_A\,\mu_A\,\dot{n}_A + V\,\dot{q} -\,\mathbf{E}\cdot\mathbf{\dot{P}} -\,\mathbf{B}\cdot\mathbf{\dot{M}}\ ,
\end{equation}
where the temperature $T$, the chemical potential $\mu_A$ of the substance $A$, the electric potential $V$, the opposite of the electric field $-\,\mathbf{E}$ and the opposite of the magnetic induction field $-\,\mathbf{B}$ are defined as the intensive conjugate fields of the extensive state fields $s$, $n_A$, $q$, $\mathbf{P}$ and $\mathbf{M}$ respectively~\cite{Callen:1960,Stratton:1941}, i.e. 
\begin{align}\label{intensive scalar}
\begin{split}
&T\equiv\frac{\partial u}{\partial s}\ ,\qquad\mu_A\equiv\frac{\partial u}{\partial n_A}\ ,\qquad V\equiv\frac{\partial u}{\partial q}\ ,\\
&-\,\mathbf{E}\equiv\frac{\partial u}{\partial \mathbf{P}}\ ,\qquad-\,\mathbf{B}\equiv\frac{\partial u}{\partial \mathbf{M}}\ .
\end{split}
\end{align}
Note that these intensive fields can only be defined locally if the system satisfies the local equilibrium hypothesis.

Using the relation~\eqref{intrinsic angular mass rest cond}, the continuity equations for the entropy~\eqref{cont eq S}, the density of the chemical substance $A$~\eqref{cont eq A}, the electric charge density~\eqref{cont eq q}, the electric polarisation~\eqref{cont eq pol} and the magnetisation~\eqref{cont eq mag}, the continuity equation for the internal energy~\eqref{cont eq int energy} is recast as,
\begin{align}\label{u balance 1}
&T\Big(\rho_s -\,\left(\bm{\nabla}\cdot\mathbf{v}\right)s -\,\bm{\nabla}\cdot\mathbf{j}_s\Big)\nonumber\\
&+ \sum_A\,\mu_A\Big(\sum_a\,\nu_{aA}\,\omega_{a} -\,\left(\bm{\nabla}\cdot\mathbf{v}\right)n_A -\,\bm{\nabla}\cdot\mathbf{j}_A\Big)\nonumber\\
&+ V\Big(-\,\left(\bm{\nabla}\cdot\mathbf{v}\right)q -\,\bm{\nabla}\cdot\mathbf{j}_q\Big)\nonumber\\
&-\,\mathbf{E}\cdot\Big(\sum_A\,\bm{\Omega}_A^{\,\text{m}}\times\mathbf{p}_A-\,\left(\bm{\nabla}\cdot\mathbf{v}\right)\mathbf{P} -\,\bm{\nabla}\cdot\mathsf{j}_{\mathbf{P}}\Big)\\
&-\,\mathbf{B}\cdot\Big(\sum_A\big(\gamma_A\,n_A\left(\mathbf{m}_A\times\mathbf{B}\right) + \bm{\Omega}_A^{\,\mathbf{M}}\times\mathbf{m}_A\big)\nonumber\\
&\phantom{-\,\mathbf{B}\cdot\ \ \ \ }-\,\left(\bm{\nabla}\cdot\mathbf{v}\right)\mathbf{M} -\,\bm{\nabla}\cdot\mathsf{j}_{\mathbf{M}}\Big)\vphantom{\sum_A}\nonumber\\
&+ \left(\bm{\nabla}\cdot\mathbf{v}\right)u + \bm{\nabla}\cdot\mathbf{j}_u = \bm{\sigma}\cdot\left(\bm{\nabla}\odot\mathbf{v}\right) + \bm{\Theta}\cdot\left(\bm{\nabla}\odot\bm{\omega}\right)\nonumber
\end{align}
Using the vectorial identities,
\begin{align*}
&T\left(\bm{\nabla}\cdot\mathbf{j}_s\right) = \bm{\nabla}\cdot\left(T\,\mathbf{j}_s\right) -\,\mathbf{j}_s\cdot\bm{\nabla}\,T\ ,\nonumber\\
&\mu_A\left(\bm{\nabla}\cdot\mathbf{j}_A\right) = \bm{\nabla}\cdot\left(\mu_A\,\mathbf{j}_A\right) -\,\mathbf{j}_A\cdot\bm{\nabla}\mu_A\ ,\nonumber\\
&V\left(\bm{\nabla}\cdot\mathbf{j}_q\right) = \bm{\nabla}\cdot\left(\mathbf{j}_q\,V\right) -\,\mathbf{j}_q\cdot\bm{\nabla}\,V\ ,\nonumber\\
&\mathbf{E}\cdot\left(\bm{\nabla}\cdot\mathsf{j}_{\mathbf{P}}\right) = \bm{\nabla}\cdot\left(\mathsf{j}_{\mathbf{P}}\cdot\mathbf{E}\right) -\,\mathsf{j}_{\mathbf{P}}\cdot\left(\bm{\nabla}\odot\mathbf{E}\right)\ ,\nonumber\\
&\mathbf{B}\cdot\left(\bm{\nabla}\cdot\mathsf{j}_{\mathbf{M}}\right) = \bm{\nabla}\cdot\left(\mathsf{j}_{\mathbf{M}}\cdot\mathbf{B}\right) -\,\mathsf{j}_{\mathbf{M}}\cdot\left(\bm{\nabla}\odot\mathbf{B}\right)\ ,\nonumber\\
&\mathbf{E}\cdot\left(\bm{\Omega}_A^{\,\text{m}}\times\mathbf{p}_A\right) = \bm{\Omega}_A^{\,\text{m}}\cdot\left(\mathbf{p}_A\times\mathbf{E}\right)\ ,\nonumber\\
&\mathbf{B}\cdot\big(\gamma_A\,n_A\left(\mathbf{m}_A\times\mathbf{B}\right)\big) = 0\ ,\nonumber\\
&\mathbf{B}\cdot\left(\bm{\Omega}_A^{\,\mathbf{M}}\times\mathbf{m}_A\right) = \bm{\Omega}_A^{\,\mathbf{M}}\cdot\left(\mathbf{m}_A\times\mathbf{B}\right)\ ,\nonumber
\end{align*}
the splitting~\eqref{sigma} of the stress tensor into a reversible and an irreversible part, and expressing the chemical irreversibility in terms of the chemical affinities $\mathcal{A}_a$, i.e.
\begin{equation}\label{chemical affinity}
\mathcal{A}_a = \sum_A\,\nu_{aA}\left(-\,\mu_A\right)\ ,
\end{equation}
the internal energy balance equation~\eqref{u balance 1} is recast as,
\begin{align}\label{u balance 2}
&\left(u - Ts + P - \sum_A\mu_A n_A - q\,V + \mathbf{P}\cdot\mathbf{E} + \mathbf{M}\cdot\mathbf{B}\right)\left(\bm{\nabla}\!\cdot\!\mathbf{v}\right)\nonumber\\
&+ \bm{\nabla}\cdot\left(\mathbf{j}_u -\,T\,\mathbf{j}_s -\,\sum_A\,\mu_A\,\mathbf{j}_A -\,\mathbf{j}_q\,V + \mathsf{j}_{\mathbf{P}}\cdot\mathbf{E} + \mathsf{j}_{\mathbf{M}}\cdot\mathbf{B}\right)\nonumber\\
&+ T\,\rho_s -\sum_a\,\omega_a\,\mathcal{A}_a + \mathbf{j}_s\cdot\bm{\nabla}\,T + \sum_A\,\mathbf{j}_A\cdot\bm{\nabla}\mu_A -\,\mathbf{j}_q\,\bm{\nabla}\,V\nonumber\\
&-\,\bm{\tilde{\sigma}}\cdot\left(\bm{\nabla}\odot\mathbf{v}\right) -\,\bm{\Theta}\cdot\left(\bm{\nabla}\odot\bm{\omega}\right)\vphantom{\sum_A}\\
&-\,\mathsf{j}_{\mathbf{P}}\cdot\left(\bm{\nabla}\odot\mathbf{E}\right) -\,\mathsf{j}_{\mathbf{M}}\cdot\left(\bm{\nabla}\odot\mathbf{B}\right)\vphantom{\sum_A}\nonumber\\
&-\,\sum_A\,\bm{\Omega}_A^{\,\text{m}}\cdot\left(\mathbf{p}_A\times\mathbf{E}\right) -\,\sum_A\,\bm{\Omega}_A^{\,\mathbf{M}}\cdot\left(\mathbf{m}_A\times\mathbf{B}\right)\vphantom{\sum_A} = 0\ .\vphantom{\sum_A}\nonumber
\end{align}
Using the expressions~\eqref{electric charge density},~\eqref{electric polarisation} and~\eqref{magnetisation} for the chemical composition of the electric charge density, the electric polarisation and magnetisation, we deduce the identities,
\begin{equation}\label{chem id 1}
\begin{split}
&q\,V = \sum_A\,q_A\,V\,n_A\ ,\\	
&\mathbf{P}\cdot\mathbf{E} = \sum_A\left(\mathbf{p}_A\cdot\mathbf{E}\right)\,n_A\ ,\\	
&\mathbf{M}\cdot\mathbf{B} = \sum_A\left(\mathbf{m}_A\cdot\mathbf{B}\right)\,n_A\ .
\end{split}
\end{equation}
Moreover, using the expressions~\eqref{el charge cur dens},~\eqref{el pol cur dens} and~\eqref{mag cur dens} for the diffusive electric current density vector, the electric polarisation and the magnetisation current tensors respectively, we obtain the identities,
\begin{equation}\label{chem id 2}
\begin{split}
&\mathbf{j}_q\,V = \sum_A\,q_A\,V\,\mathbf{j}_A\ ,\\	
&\mathsf{j}_{\mathbf{P}}\cdot\mathbf{E} = \sum_A\left(\mathbf{p}_A\cdot\mathbf{E}\right)\,\mathbf{j}_A\ ,\\	
&\mathsf{j}_{\mathbf{M}}\cdot\mathbf{B} = \sum_A\left(\mathbf{m}_A\cdot\mathbf{B}\right)\,\mathbf{j}_A\ .
\end{split}
\end{equation}
Furthermore, using the expressions~\eqref{mom cur dens},~\eqref{ang mom cur dens},~\eqref{el pol cur dens} and~\eqref{mag cur dens}, for the momentum and intrinsic angular momentum stress tensors as well as for the electric polarisation and the magnetisation current tensors respectively, we obtain the identities,
\begin{equation}\label{chem id 3}
\begin{split}
&\bm{\tilde{\sigma}}\cdot\left(\bm{\nabla}\odot\mathbf{v}\right) = \sum_A\,\mathbf{j}_A\cdot\Big(-\,m_A\,\mathbf{v}_A\,\bm{\nabla}\,\mathbf{v}\Big)\ ,\\	
&\bm{\Theta}\cdot\left(\bm{\nabla}\odot\bm{\omega}\right) = \sum_A\,\mathbf{j}_A\cdot\Big(-\,I_A\,\bm{\omega}_A\,\bm{\nabla}\,\bm{\omega}\Big)\ ,\\
&\mathsf{j}_{\mathbf{P}}\cdot\left(\bm{\nabla}\odot\mathbf{E}\right) = \sum_A\,\mathbf{j}_A\cdot\Big(\mathbf{p}_A\,\bm{\nabla}\,\mathbf{E}\Big)\ ,\\	
&\mathsf{j}_{\mathbf{M}}\cdot\left(\bm{\nabla}\odot\mathbf{B}\right) = \sum_A\,\mathbf{j}_A\cdot\Big(\mathbf{m}_A\,\bm{\nabla}\,\mathbf{B}\Big)\ .
\end{split}
\end{equation}
At this point it is useful to introduce the generalised chemical potential $\bar{\mu}_A$ that is defined as,
\begin{equation}\label{chem pot}
\bar{\mu}_A = \mu_A + q_A\,V -\,\mathbf{p}_A\cdot\mathbf{E} -\,\mathbf{m}_A\cdot\mathbf{B}\ ,
\end{equation}

Using the identities~\eqref{chem id 1}-\eqref{chem id 3} and the definition~\eqref{chem pot}, the internal energy balance equation~\eqref{u balance 2} is recast explicitly in terms of the physical properties of the chemical components, i.e.
\begin{align}\label{u balance 3}
&\Bigg(u -\,T\,s + P - \sum_A\,\bar{\mu}_A\,n_A\Bigg)\left(\bm{\nabla}\cdot\mathbf{v}\right)\nonumber\\
&+ \bm{\nabla}\cdot\left(\mathbf{j}_u -\,T\,\mathbf{j}_s -\,\sum_A\,\bar{\mu}_A\,\mathbf{j}_A\right)\nonumber\\
&+ T\,\rho_s -\,\sum_a\,\omega_a\,\mathcal{A}_a -\,\mathbf{j}_s\cdot\left(-\,\bm{\nabla}\,T\right)\nonumber\\
&-\,\sum_A\,\mathbf{j}_A\cdot\Big(-\,\bm{\nabla}\mu_A -\,q_A\,\bm{\nabla}\,V\Big)\\
&-\,\sum_A\,\mathbf{j}_A\cdot\Big(-\,m_A\,\mathbf{v}_A\,\bm{\nabla}\,\mathbf{v} -\,I_A\,\bm{\omega}_A\,\bm{\nabla}\,\bm{\omega}\Big)\nonumber\\
&-\,\sum_A\,\mathbf{j}_A\cdot\Big(\mathbf{p}_A\,\bm{\nabla}\,\mathbf{E} + \mathbf{m}_A\,\bm{\nabla}\,\mathbf{B}\Big)\nonumber\\
&-\,\sum_A\,\bm{\Omega}_A^{\,\text{m}}\cdot\left(\mathbf{p}_A\times\mathbf{E}\right) -\,\sum_A\,\bm{\Omega}_A^{\,\mathbf{M}}\cdot\left(\mathbf{m}_A\times\mathbf{B}\right) = 0\vphantom{\sum_A}\nonumber\ .
\end{align}

Firstly, the internal energy balance equation~\eqref{u balance 3} has to hold locally for all flows. This implies that the terms in the first brackets have to vanish, which yields the thermostatic equilibrium equation for the continuous medium, i.e.
\begin{equation}\label{thermostatic eq}
u = T\,s - P + \sum_A\,\Big(\mu_A + q_A\,V -\,\mathbf{p}_A\cdot\mathbf{E} -\,\mathbf{m}_A\cdot\mathbf{B}\Big)\,n_A\ .
\end{equation}

Secondly, the internal energy balance equation~\eqref{u balance 3} has to hold locally for all currents. This implies that the terms in the second brackets have to vanish, which yields the reversible thermodynamic evolution equation for the continuous medium, i.e.
\begin{equation}\label{rev thermodyn eq}
\mathbf{j}_u = T\,\mathbf{j}_s + \sum_A\,\Big(\mu_A + q_A\,V -\,\mathbf{p}_A\cdot\mathbf{E} -\,\mathbf{m}_A\cdot\mathbf{B}\Big)\,\mathbf{j}_A\ .
\end{equation}

Thirdly, the thermostatic equilibrium equation~\eqref{thermostatic eq} and the reversible thermodynamic evolution equation~\eqref{rev thermodyn eq} imply that the internal energy balance equation~\eqref{u balance 3} yields the irreversible thermodynamic evolution equation for the continuous medium, i.e.
\begin{align}\label{irrev thermodyn equ}
&\rho_s=\frac{1}{T}\left\{\sum_a\,\omega_a\,\mathcal{A}_a + \mathbf{j}_s\cdot\left(-\,\bm{\nabla}\,T\right)\right.\nonumber\\
&\phantom{\rho_s=\frac{1}{T}\ \ \ } + \sum_A\,\mathbf{j}_A\cdot\Big(-\,\bm{\nabla}\mu_A -\,q_A\,\bm{\nabla}\,V\Big)\nonumber\\
&\phantom{\rho_s=\frac{1}{T}\ \ \ } + \sum_A\,\mathbf{j}_A\cdot\Big(-\,m_A\,\mathbf{v}_A\,\bm{\nabla}\,\mathbf{v} -\,I_A\,\bm{\omega}_A\,\bm{\nabla}\,\bm{\omega}\Big)\nonumber\\
&\phantom{\rho_s=\frac{1}{T}\ \ \ } + \sum_A\,\mathbf{j}_A\cdot\Big(\mathbf{p}_A\,\bm{\nabla}\,\mathbf{E} + \mathbf{m}_A\,\bm{\nabla}\,\mathbf{B} \Big)\\
&\phantom{\rho_s=\frac{1}{T}\ \ \ } \left. + \sum_A\,\bm{\Omega}_A^{\,\text{m}}\cdot\left(\mathbf{p}_A\times\mathbf{E}\right) + \sum_A\,\bm{\Omega}_A^{\,\mathbf{M}}\cdot\left(\mathbf{m}_A\times\mathbf{B}\right)\right\}\ .\nonumber
\end{align}

Finally, note that the time derivative of the thermostatic equation~\eqref{thermostatic eq} and the expression~\eqref{deriv u} for the time derivative of the internal energy determines a generalised Gibbs-Duhem relation that yields the time evolution of the intensive fields~\eqref{intensive scalar}, i.e.
\begin{equation}\label{Gibbs Duhem}
s\,\dot{T} -\,\dot{P} + \sum_A\,n_A\,\dot{\mu}_A + q\,\dot{V} -\,\mathbf{P}\cdot\mathbf{\dot{E}} -\,\mathbf{M}\cdot\mathbf{\dot{B}} = 0\ .
\end{equation}
%


\section{Thermodynamical phenomenology}
\label{Thermodynamical phenomenology}

\subsection{Linear phenomenological relations and Onsager matrix elements}

In order to deduce the linear phenomenological relations, we extend the approach developed by Onsager~\cite{Onsager:1931a,Onsager:1931b} to include intrinsic rotations. The expression~\eqref{irrev thermodyn equ} can be formally split into a scalar, a vectorial and a pseudo-vectorial part, which are irreducible representations of the Euclidean group and have different symmetries. Thus, the entropy source density~\eqref{irrev thermodyn equ} is expressed formally as,
\begin{equation}\label{Onsager i}
\rho_s=\frac{1}{T}\left\{\sum_a\,\omega_a\,\mathcal{A}_a + \sum_{\alpha}\,\mathbf{j}_{\alpha}\cdot\mathbf{F}_{\alpha} + \sum_{\substack{A,\,i}}\,\bm{\Omega}_{A}^{\,i}\cdot\mathbf{T}_{A}^{\,i}\right\}\ ,
\end{equation}
where $\omega_a$ is a scalar chemical reaction rate density, $\mathcal{A}_a$ is a scalar chemical affinity, $\mathbf{j}_{\alpha}$ is a vectorial current density, $\mathbf{F}_{\alpha}$ is a vectorial force, $\bm{\Omega}_{A}^{\,i}$ is a pseudo-vectorial intrinsic rotation rate density and $\mathbf{T}_{A}^{\,i}$ is a pseudo-vectorial intrinsic torque.

In the relation~\eqref{Onsager i}, there are two types of vectorial current densities and forces (i.e. $\alpha\in\{\vphantom{a}_s,\vphantom{a}_A\}$). First, there is a thermal current density $\mathbf{j}_s$ and an irreversible thermal force $\mathbf{F}_s=-\,\bm{\nabla}\,T$. Second, there are current densities $\mathbf{j}_A$ for the chemical substances $A$ and irreversible forces $\mathbf{F}_A$ acting on these substances. The expression for the forces $\mathbf{F}_A$ is given by the irreversible relation~\eqref{irrev thermodyn equ}, i.e.
\begin{align}\label{force}	
&\mathbf{F}_A = -\,\bm{\nabla}\mu_A -\,q_A\,\bm{\nabla}\,V -\,m_A\,\mathbf{v}_A\,\bm{\nabla}\,\mathbf{v} -\,I_A\,\bm{\omega}_A\,\bm{\nabla}\,\bm{\omega}\nonumber\\ 
&\phantom{\mathbf{F}_A = } + \mathbf{p}_A\,\bm{\nabla}\,\mathbf{E} + \mathbf{m}_A\,\bm{\nabla}\,\mathbf{B}\ ,
\end{align}
where the first term on the RHS is the chemical force, the second term is the Coulomb force, the third and the fourth terms are the viscous forces, the fifth term is the electric polarisation force and the sixth term is the magnetisation force. Note that in a stationary regime, the viscous forces, the electric polarisation force and the magnetisation force generate dielectrophoresis and magnetophoresis.

In the relation~\eqref{Onsager i}, there are also two types of pseudo-vectorial intrinsic rotation rate densities and torques (i.e. $\vphantom{a}_A^i\in\{\vphantom{a}_A^\text{m},\vphantom{a}_A^\mathbf{M}\}$) associated to the chemical substances $A$. First, there are intrinsic rotation rate densities $\bm{\Omega}_A^{\,\text{m}}$ and irreversible torques $\mathbf{T}_{A}^{\,\text{m}}$ associated with the intrinsic rotational motion of the matter. Second, there are intrinsic rotation rate densities $\bm{\Omega}_A^{\,\mathbf{M}}$ and irreversible torques $\mathbf{T}_{A}^{\,\mathbf{M}}$ associated with micro-magnetism and matter. The expression for the torques $\mathbf{T}_{A}^{\,i}$ is given by the irreversible relation~\eqref{irrev thermodyn equ}, i.e. 
\begin{equation}\label{torque}	
\begin{split}
&\mathbf{T}_A^{\,\text{m}} = \mathbf{p}_A\times\mathbf{E}\ ,\\
&\mathbf{T}_A^{\,\mathbf{M}} = \mathbf{m}_A\times\mathbf{B}\ .
\end{split}
\end{equation}

The local expression of the second law~\eqref{cont eq S} requires the entropy source density to be locally positive definite, i.e. $\rho_s\geqslant 0$. Extending Onsager's approach, in the neighbourhood of a local thermodynamic state where the scalar affinities $\mathcal{A}_a$, the vectorial forces $\mathbf{F}_{\alpha}$ and the pseudo-vectorial intrinsic torques $\mathbf{T}_{A}^i$ are sufficiently small, the entropy source density can be expressed formally as the sum of quadratic forms of $\mathcal{A}_a$, $\mathbf{F}_{\alpha}$ and $\mathbf{T}_{A}^i$, which ensures that it is positive definite, i.e. 
\begin{equation}\label{Onsager quad}
\begin{split}
&\rho_s=\frac{1}{T}\Big(\sum_{a,\,b}\,L_{ab}\,\mathcal{A}_{a}\,\mathcal{A}_{b} + \sum_{\alpha,\,\beta}\,\mathsf{L}_{\alpha\beta}\cdot\left(\mathbf{F}_{\alpha}\odot\mathbf{F}_{\beta}\right)\\	
&\phantom{\rho_s=\frac{1}{T}\Big(} + \sum_{\substack{A,\,B\\i,\,j}}\,\mathsf{L}_{AB}^{\,ij}\cdot\left(\mathbf{T}_A^{\,i}\odot\mathbf{T}_B^{\,j}\right)\Big)\geqslant 0\ ,
\end{split}
\end{equation}
where the phenomenological components are the Onsager matrix elements, which are of two different types: scalars $L_{\alpha\beta}$ and rank-$2$ tensors $\mathsf{L}_{\alpha\beta}$ and $\mathsf{L}_{AB}^{\,ij}$. The symmetries of the Onsager matrices are given by the Onsager reciprocity relations, i.e.
\begin{equation}\label{Onsager reciprocity relations}
\begin{split}
&L_{ab}\left(s,n_A,q,\mathbf{P},\mathbf{M}\right) = L_{ba}\left(s,n_A,q,\mathbf{P},-\,\mathbf{M}\right)\ ,\\	
&\mathsf{L}_{\alpha\beta}\left(s,n_A,q,\mathbf{P},\mathbf{M}\right) = \mathsf{L}_{\beta\alpha}\left(s,n_A,q,\mathbf{P},-\,\mathbf{M}\right)\ ,\\
&\mathsf{L}_{AB}^{\,ij}\left(s,n_A,q,\mathbf{P},\mathbf{M}\right) = \mathsf{L}_{AB}^{\,ji}\left(s,n_A,q,\mathbf{P},-\,\mathbf{M}\right)\ ,
\end{split}
\end{equation}
which cannot be derived within a thermodynamic approach but require a statistical treatment since they are a consequence of the time reversibility of the microscopic dynamics~\cite{Onsager:1931a}. The inequality~\eqref{Onsager quad} has to hold for each part, which implies that each quadratic form has to be positive definite. 

Thus, the chemical reaction rate densities $\omega_a$ are related to the chemical affinities $\mathcal{A}_{b}$ through scalar linear phenomenological relations, i.e.
\begin{equation}
\label{Onsager scalars}
\omega_a = \displaystyle{\sum_{b}}\,L_{ab}\,\mathcal{A}_b\ ,
\end{equation}
where the Onsager matrix has to satisfy,
\begin{equation}\label{scalar Onsager}
\frac{1}{T}\left\{L_{ab}\right\}\geqslant 0\ ,
\end{equation}
in order for the scalar quadratic form in the relation~\eqref{Onsager quad} to be positive definite. The scalar linear phenomenological relations~\eqref{Onsager scalars} account for the irreversibility due to the chemistry. Note that these linear relations are only accurate provided that the chemical affinities $\mathcal{A}_b$ are sufficiently small compared to the thermal excitation, i.e. $\mathcal{A}_b\ll k_B T$ where $k_B$ is Boltzmann's constant, as pointed out by Glansdorf and Prigogine~\cite{Glansdorf:1971} and de Groot and Mazur~\cite{deGroot:1984}.

Similarly, the vectorial current densities $\mathbf{j}_{\alpha}$ are related to the vectorial forces $\mathbf{F}_{\beta}$ through vectorial linear phenomenological relations, i.e.
\begin{equation}\label{Onsager reciprocity rel vec}
\mathbf{j}_{\alpha}=\sum_{\beta}\,\mathsf{L}_{\alpha\beta}\cdot\mathbf{F}_{\beta}\ ,
\end{equation}
where the Onsager matrix has to satisfy,
\begin{equation}\label{Onsager quad cond vec}
\frac{1}{T}\left\{\mathsf{L}_{\alpha\beta}\right\}\geqslant \mathsf{0}\ ,	
\end{equation}
in order for the vectorial quadratic form in the relation~\eqref{Onsager quad} to be positive definite. The vectorial linear phenomenological relations~\eqref{Onsager reciprocity rel vec} are expressed explicitly in terms of the currents densities $\mathbf{j}_s$ and $\mathbf{j}_A$ and forces $\mathbf{F}_s$ and $\mathbf{F}_A$ as,
\begin{equation}
\label{Onsager vectors}
\begin{cases}
\mathbf{j}_s = \mathsf{L}_{ss}\cdot\left(-\,\bm{\nabla}\,T\right) +  \displaystyle{\sum_{B}}\,\mathsf{L}_{sB}\cdot\mathbf{F}_B\\
\mathbf{j}_A = \mathsf{L}_{As}\cdot\left(-\,\bm{\nabla}\,T\right) +  \displaystyle{\sum_{B}}\,\mathsf{L}_{AB}\cdot\mathbf{F}_B
\end{cases}
\end{equation}
where the Onsager matrix~\eqref{Onsager quad cond vec} is positive definite, i.e.
\begin{equation}\label{vector Onsager}
\frac{1}{T}\,
\begin{pmatrix} 
\mathsf{L}_{ss} & \mathsf{L}_{sB} \vphantom{\displaystyle{\frac{x}{x}}}\\ 
\mathsf{L}_{As} & \mathsf{L}_{AB} \vphantom{\displaystyle{\frac{x}{x}}}
\end{pmatrix}
\geqslant \mathsf{0}\ .
\end{equation}
The vectorial linear phenomenological relations~\eqref{Onsager vectors} account for the irreversibility due to the transport.

Likewise, the pseudo-vectorial intrinsic rotation rate densities $\bm{\Omega}_A^{\,i}$ are related to the pseudo-vectorial torques $\mathbf{T}_B^{\,j}$ through pseudo-vectorial linear phenomenological relations, i.e.
\begin{equation}\label{Onsager reciprocity rel pseudo-vec}
\bm{\Omega}_A^{\,i}=\sum_{B,\,j}\,\mathsf{L}_{AB}^{\,ij}\cdot\mathbf{T}_B^{\,j}\ ,
\end{equation}
where the Onsager matrix has to satisfy,
\begin{equation}\label{Onsager quad cond pseudo-vec}
\frac{1}{T}\left\{\mathsf{L}_{AB}^{\,ij}\right\}\geqslant \mathsf{0}\ ,	
\end{equation}
in order for the pseudo-vectorial quadratic form in the relation~\eqref{Onsager quad} to be positive definite. The pseudo-vectorial linear phenomenological relations~\eqref{Onsager reciprocity rel pseudo-vec} are expressed explicitly in terms of the intrinsic rotation rate densities $\bm{\Omega}_A^{\,i}$ and torques $\mathbf{T}_B^{\,j}$ as,
\begin{equation}
\label{Onsager pseudo-vectors}
\begin{cases}
\bm{\Omega}_A^{\,\text{m}} = \displaystyle{\sum_{B}}\Big(\mathsf{L}_{AB}^{\text{m}\text{m}}\cdot\left(\mathbf{p}_B\times\mathbf{E}\right) + \mathsf{L}_{AB}^{\text{m}\mathbf{M}}\cdot\left(\mathbf{m}_B\times\mathbf{B}\right)\Big)\\
\bm{\Omega}_A^{\,\mathbf{M}} = \displaystyle{\sum_{B}}\Big(\mathsf{L}_{AB}^{\mathbf{M}\text{m}}\cdot\left(\mathbf{p}_B\times\mathbf{E}\right) + \mathsf{L}_{AB}^{\mathbf{M}\mathbf{M}}\cdot\left(\mathbf{m}_B\times\mathbf{B}\right)\Big)
\end{cases}
\end{equation}
where the Onsager matrix~\eqref{Onsager quad cond pseudo-vec} is positive definite, i.e.
\begin{equation}\label{pseudo-vector Onsager}
\frac{1}{T}\,
\begin{pmatrix} 
\mathsf{L}_{AB}^{\text{m}\text{m}} & \mathsf{L}_{AB}^{\text{m}\mathbf{M}} \vphantom{\displaystyle{\frac{x}{x}}}\\ 
\mathsf{L}_{AB}^{\mathbf{M}\text{m}} & \mathsf{L}_{AB}^{\mathbf{M}\mathbf{M}} \vphantom{\displaystyle{\frac{x}{x}}}
\end{pmatrix}
\geqslant \mathsf{0}\ .
\end{equation}
The pseudo-vectorial linear phenomenological relations~\eqref{Onsager pseudo-vectors} account for the irreversibility due to the relaxation. They also represent the classical counterpart of the spin-orbit coupling, since they couple the rotation rate density $\bm{\Omega}_A^{\,\mathbf{M}}$ of the magnetisation with the rotation rate density of the matter $\bm{\Omega}_A^{\,\text{m}}$.

It is worth emphasising that although the scalar~\eqref{Onsager scalars}, the vectorial~\eqref{Onsager vectors} and the pseudo-vectorial~\eqref{Onsager pseudo-vectors} linear phenomenological relations are structurally independent, they are coupled through the time evolution equations. The chemical affinities $\mathcal{A}_a$ defined in relation~\eqref{chemical affinity} couple the scalar~\eqref{Onsager scalars} and vectorial~\eqref{Onsager vectors} linear phenomenological relations. This coupling has interesting applications for spintronics, some of which were investigated in~\cite{Ansermet:2008}.

In the particular case of a continuous medium made of a single chemical substance $A$, the scalar linear relations~\eqref{Onsager scalars} vanish, the vectorial linear relations~\eqref{Onsager vectors} become,
\begin{equation}\label{Onsager vectors bis}
\begin{cases}
\mathbf{j}_s = \mathsf{L}_{ss}\cdot\left(-\,\bm{\nabla}\,T\right) +  \mathsf{L}_{sA}\cdot\mathbf{F}_A   \vphantom{\displaystyle{\frac{x}{x}}}\\
\mathbf{j}_A = \mathsf{L}_{As}\cdot\left(-\,\bm{\nabla}\,T\right) +  \mathsf{L}_{AA}\cdot\mathbf{F}_A   \vphantom{\displaystyle{\frac{x}{x}}}
\end{cases}
\end{equation}
and the pseudo-vectorial linear relations~\eqref{Onsager pseudo-vectors} reduce to,
\begin{equation}
\label{Onsager pseudo-vectors bis}
\begin{cases}
\bm{\Omega}_A^{\,\text{m}} = \mathsf{L}_{AA}^{\text{m}\text{m}}\cdot\left(\mathbf{p}_A\times\mathbf{E}\right) + \mathsf{L}_{AA}^{\text{m}\mathbf{M}}\cdot\left(\mathbf{m}_A\times\mathbf{B}\right)   \vphantom{\displaystyle{\frac{x}{x}}}\\
\bm{\Omega}_A^{\,\mathbf{M}} = \mathsf{L}_{AA}^{\mathbf{M}\text{m}}\cdot\left(\mathbf{p}_A\times\mathbf{E}\right) + \mathsf{L}_{AA}^{\mathbf{M}\mathbf{M}}\cdot\left(\mathbf{m}_A\times\mathbf{B}\right)  \vphantom{\displaystyle{\frac{x}{x}}}
\end{cases}
\end{equation}

\subsection{Physical applications}

\subsubsection{Lehmann effect}

\label{Lehmann effect}

In $1900$, Lehmann~\cite{Lehmann:1900} observed that droplets of liquid crystals, that have a chiral cholesteric structure, were set in rotation by a temperature gradient. The explanation given by Leslie~\cite{Leslie:1968} is now questioned~\cite{Oswald:2012} by recent observations. Here, we show how our formalism accounts for the Lehmann effect.

It is worth mentioning that the irreversible thermodynamics of a continuum of nematic liquid crystals has been examined by Müller~\cite{Mueller:1985} and Muschik~\cite{Muschik:1991,Muschik:2004}, even though these authors did not explicitly deduce the Lehmann effect from their formalism. The ``director'' vector field they introduced corresponds to the preferred axis $\mathbf{\hat{n}}$ of the continuum of liquid crystals. In contrast to them, we do not enlarge the spatial manifold to account for $\mathbf{\hat{n}}$. Instead, we require the net electric polarisation $\mathbf{P}$, which is a state field, to be collinear to $\mathbf{\hat{n}}$.

We consider a uniform continuum made of identical liquid crystals that are oriented in the same direction. The continuum of liquid crystals is gyrotropic since the chirality of the liquid crystals defines a preferred axis of unit vector $\mathbf{\hat{n}}$. Each liquid crystal is helicoidal and made of identical elements $A$ that are dielectrics, i.e. $q_A=0$, and have electric dipoles $\mathbf{p}_A$ orthogonal to the helix axis~\cite{Kim:1982}. Thus, in the absence of an external interaction, the liquid crystals have no net electric polarisation. On a macroscopic scale, the continuum is homogeneous, i.e. $\bm{\nabla}\,\mu_A = \mathbf{0}$. We assume that the viscosity can be neglected, i.e. $\bm{\nabla}\,\mathbf{v}_A = \mathbf{0}$ and $\bm{\nabla}\,\bm{\omega}_A = \mathbf{0}$. The liquid crystals are trapped in an experimental set-up such that they have no translational motion, i.e. $\mathbf{j}_A = \mathbf{0}$. Moreover, a temperature gradient $\bm{\nabla}\,T$ is applied along a direction that is different from the preferred axis of the crystals. 

In such a case, the linear phenomenological relation~\eqref{Onsager vectors bis} reduces to,
\begin{equation}\label{Seebeck effect 00}
\mathbf{p}_A\,\bm{\nabla}\,\mathbf{E} = \left(\mathsf{L}_{AA}^{-1}\cdot\mathsf{L}_{As}\right)\cdot\bm{\nabla}\,T\ ,
\end{equation}
where the temperature gradient $\bm{\nabla}\,T$ induces a electric field $\mathbf{E}$ and the electric dipoles $\mathbf{p}_A$ rotate in an asymmetric manner in order to lower the Debye energy $-\,\mathbf{p}_A\cdot\mathbf{E}$. Thus, a net electric polarisation $\mathbf{P}$ is generated along the preferred axis $\mathbf{\hat{n}}$, i.e.
\begin{equation}\label{pol axis}
\mathbf{P} = n_A\,\mathbf{p}_A = n_A\left(\mathbf{p}_A\cdot\mathbf{\hat{n}}\right)\mathbf{\hat{n}}\ ,
\end{equation}
and the linear relation~\eqref{Seebeck effect 00} is recast as,
\begin{equation}\label{Seebeck effect 0}
\mathbf{P}\,\bm{\nabla}\,\mathbf{E} = \left(n_A\,\mathsf{L}_{AA}^{-1}\cdot\mathsf{L}_{As}\right)\cdot\bm{\nabla}\,T\ .
\end{equation}
The spatial symmetry requires the electric polarisation force $\mathbf{P}\,\bm{\nabla}\,\mathbf{E}$ and the thermal force $-\,\bm{\nabla}\,T$ to be collinear, i.e.
\begin{equation}\label{Seebeck effect 0 bis}
\mathbf{P}\,\bm{\nabla}\,\mathbf{E} = n_A\,L_{AA}^{-1}\,L_{As}\,\bm{\nabla}\,T\ ,
\end{equation}
where $\mathsf{L}_{AA}^{-1} = L_{AA}^{-1}\,\mathbb{1}$ and $\mathsf{L}_{As} = L_{As}\,\mathbb{1}$. Thus, the phenomenological relation~\eqref{Seebeck effect 0 bis} can be recast as,
\begin{equation}\label{Seebeck effect 0 ter}
\mathbf{P}\,\bm{\nabla}\,\mathbf{E} = \lambda\,n_A\,k_B\,\bm{\nabla}\,T\ ,
\end{equation}
where $\lambda$ is a dimensionless parameter given by,
\begin{equation}\label{en equipartition}
\lambda = \frac{L_{AA}^{-1}\,L_{As}}{k_B}\ .
\end{equation}
In the absence of an applied magnetic field, i.e. $\mathbf{B} = \mathbf{0}$, the electric field $\mathbf{E}$ is expressed in terms of the electric potential $V$ as,
\begin{equation}\label{Faraday bis}
\mathbf{E} = -\,\bm{\nabla}\,V\ .
\end{equation}
The electric polarisation force density $\mathbf{P}\,\bm{\nabla}\,\mathbf{E}$ satisfies the vectorial identity,
\begin{equation}\label{rel vect Lehmann}
\begin{split}
&\mathbf{P}\,\bm{\nabla}\,\mathbf{E} = \left(\mathbf{P}\cdot\bm{\nabla}\right)\mathbf{E} + \mathbf{P}\times\left(\bm{\nabla}\times\mathbf{E}\right) = \left(\mathbf{P}\cdot\bm{\nabla}\right)\mathbf{E}\\
&\phantom{\mathbf{P}\,\bm{\nabla}\,\mathbf{E}} = \bm{\nabla}_{\mathbf{P}}\cdot\left(\mathbf{P}\odot\mathbf{E}\right) + \left(-\,\bm{\nabla}\cdot\mathbf{P}\right)\mathbf{E}\ ,
\end{split}
\end{equation}
where we used Faraday's law~\eqref{Faraday bis} and the index $\vphantom{a}_\mathbf{P}$ denotes that there is a dot product between the covariant differential operator $\bm{\nabla}$ and the electric polarisation $\mathbf{P}$. The term $\bm{\nabla}_{\mathbf{P}}\cdot\left(\mathbf{P}\odot\mathbf{E}\right)$ in the identity~\eqref{rel vect Lehmann} corresponds to a surface contribution after integration over the volume of the liquid crystal continuum~\cite{Reuse:2012}. Thus, it can be neglected in the bulk of the continuum where the electric polarisation force density $\mathbf{P}\,\bm{\nabla}\,\mathbf{E}$ is expressed as,
\begin{equation}\label{elec pol force dens}
\mathbf{P}\,\bm{\nabla}\,\mathbf{E} = \left(-\,\bm{\nabla}\cdot\mathbf{P}\right)\mathbf{E}\ .
\end{equation}
Moreover, using the definition of the bound electric charge density $q_{\mathbf{P}}$, i.e.
\begin{equation}\label{bound charge density}
q_{\mathbf{P}} \equiv -\,\bm{\nabla}\cdot\mathbf{P}\ ,
\end{equation}
the electric polarisation force density~\eqref{elec pol force dens} is recast as~\cite{Reuse:2012},
\begin{equation}\label{rel vect Lehmann bis}
\mathbf{P}\,\bm{\nabla}\,\mathbf{E} = q_{\mathbf{P}}\,\mathbf{E}\ ,
\end{equation}
which shows that it is the analog of the electric part of the Lorentz force density for bound charges. Substituting the expression~\eqref{rel vect Lehmann bis} in the linear relation~\eqref{Seebeck effect 0 ter}, the latter can be expressed as a Seebeck effect for bound electric charges in an electric insulator, i.e.
\begin{equation}\label{Seebeck effect}
\mathbf{E} = \varepsilon_{\mathbf{P}}\,\bm{\nabla}\,T\ ,
\end{equation}
where the Seebeck coefficient for bound electric charges,
\begin{equation*}
\varepsilon_{\mathbf{P}} \equiv \frac{\lambda\,n_A\,k_B}{q_{\mathbf{P}}}\ .
\end{equation*}
Moreover, the liquid crystals are subjected to an external electric torque~\eqref{torque} that sets them in rotational motion. Thus, the external torque density is given by, 
\begin{equation}\label{cond Lehmann 00}
\bm{\tau}^{\,\text{ext}} = n_A\,\mathbf{T}_A^{\,\text{m}} = \mathbf{P}\times\mathbf{E}\ .
\end{equation}
Using the equation~\eqref{Seebeck effect} for the Seebeck effect, the torque density $\bm{\tau}^{\,\text{ext}}$ is recast as,
\begin{equation}\label{cond Lehmann 0}
\bm{\tau}^{\,\text{ext}} = \varepsilon_{\mathbf{P}}\left(\mathbf{P}\times\bm{\nabla}\,T\right)\ .
\end{equation}
The rotation occurs in the plane spanned by the electric polarisation $\mathbf{P}$ and the temperature gradient $\bm{\nabla}\,T$ around the local centre of mass of the liquid crystals. Taking into account the absence of a matter current, i.e. $\bm{\Theta} = \mathbf{0}$, and substituting the expression~\eqref{cond Lehmann 0} for the torque density into Newton's seond law in intrinsic rotation~\eqref{Euler eq}, the latter accounts for the Lehmann effect, i.e.
\begin{equation}\label{Euler eq Lehmann}
\bm{\dot{\omega}} = \frac{\varepsilon_{\mathbf{P}}}{I}\,\left(\mathbf{P}\times\bm{\nabla}\,T\right)\ ,
\end{equation}
where $I$ is the intrinsic angular mass density of the liquid crystals.
The intensity of this effect is expressed as
\begin{equation}\label{Euler eq Lehmann sca}
\ddot{\theta} = \frac{\varepsilon_{\mathbf{P}}\,P\,\nabla\,T}{I}\,\sin\theta\ ,
\end{equation}
where $\theta$ is the angle between the electric polarisation $\mathbf{P}$ and the temperature gradient $\bm{\nabla}\,T$, i.e. $|\mathbf{P}\times\bm{\nabla}\,T| = P\,\nabla\,T\,\sin\theta$.

\subsubsection{Electric Lehmann effect}

In $1974$, de Gennes~\cite{deGennes:1974} noticed that liquid crystals were set in rotation by an electric field. This effect is called the electric Lehmann effect~\cite{Dequidt:2007} since it corresponds to a Lehmann effect where the driving force is an electric force instead of a thermal force. Here, we show how our formalism accounts for the electric Lehmann effect.

We consider the same continuum of liquid crystals as in Sec.~\ref{Lehmann effect}. In contrast to the Lehmann effect, the electric Lehmann effect is an isothermal effect. The temperature gradient $\bm{\nabla}\,T$ is replaced by an electric potential gradient $\bm{\nabla}\,V$ that is applied along a direction that is different from the preferred axis of the liquid crystals. By analogy with the Lehmann effect, the electric field must be inhomogeneous in order to generate an electric polarisation force density~\eqref{rel vect Lehmann bis}.

In the presence of an applied electric potential gradient $\bm{\nabla}\,V$, the electric dipoles $\mathbf{p}_A$ rotate in an asymmetric manner in order to lower the Debye energy $\mathbf{p}_A\cdot\bm{\nabla}\,V$, which yields a net electric polarisation $\mathbf{P}$ along the preferred axis $\mathbf{\hat{n}}$ given in equation~\eqref{pol axis}. Moreover, the liquid crystals are subjected to an external electric torque density~\eqref{cond Lehmann 00}, that sets them in rotational motion. In the absence of a magnetic induction field $\mathbf{B}$, Faraday's law~\eqref{Faraday bis} implies that the torque density~\eqref{cond Lehmann 00} yields,
\begin{equation}\label{cond Lehmann E}
\bm{\tau}^{\,\text{ext}} = -\,\mathbf{P}\times\bm{\nabla}\,V\ .
\end{equation}
The rotation occurs in the plane spanned by the electric polarisation $\mathbf{P}$ and the electric potential gradient $\bm{\nabla}\,V$ around the local centre of mass of the liquid crystals. Taking into account the absence of a matter current, i.e. $\bm{\Theta} = \mathbf{0}$, and substituting the expression~\eqref{cond Lehmann E} for the torque density into Newton's second law in intrinsic rotation~\eqref{Euler eq}, the latter accounts for the electric Lehmann effect, i.e.
\begin{equation}\label{Euler eq Lehmann E}
\bm{\dot{\omega}} = -\,\frac{1}{I}\,\left(\mathbf{P}\times\bm{\nabla}\,V\right)\ .
\end{equation}
The intensity of this effect is expressed as
\begin{equation}\label{Euler eq Lehmann sca E}
\ddot{\theta} = -\,\frac{P\,\nabla\,V}{I}\,\sin\theta\ ,
\end{equation}
where $\theta$ is the angle between the electric polarisation $\mathbf{P}$ and the electric potential gradient $\bm{\nabla}\,V$, i.e. $|\mathbf{P}\times\bm{\nabla}\,V| = P\,\nabla\,V\,\sin\theta$.

It is useful to point out that Quincke~\cite{Quincke:1896} observed in $1896$ that small dielectric spheres in suspension in a liquid placed inside a parallel-plate capacitor are set in rotational motion by charging the capacitor. The intrinsic rotational dynamics of the Quincke effect is described by the same time evolution equations~\eqref{Euler eq Lehmann E} and~\eqref{Euler eq Lehmann sca E} as the electric Lehmann effect.

\subsubsection{Relaxation of electric dipoles}

The first model of the relaxation of electric dipoles $\mathbf{p}_A$ is due to Debye~\cite{Debye:1913}. Here, we show how our formalism accounts for this relaxation.

We consider a homogeneous sample made of a single chemical substance $A$, consisting of electric dipoles $\mathbf{p}_A$ in the absence of a magnetic induction field, i.e. $\mathbf{B} = \mathbf{0}$. In the frame of sample, the chemical substance $A$ has no translational motion, i.e. $\mathbf{j}_A=\mathbf{0}$.

The absence of a magnetic induction field, i.e. $\mathbf{B} = \mathbf{0}$, implies that the linear phenomenological relation~\eqref{Onsager pseudo-vectors bis} reduces to,
\begin{equation}\label{relaxation el pol 0}
\bm{\Omega}_A^{\,\text{m}} = \mathsf{L}_{AA}^{\text{m}\text{m}}\cdot\left(\mathbf{p}_A\times\mathbf{E}\right)\ . 
\end{equation}
The process is irreversible and dissipative, which means that the Debye energy $-\,\mathbf{p}_A\cdot\mathbf{E}$ has to diminish. This implies that the intrinsic rotation of the local element, to which the electric dipole $\mathbf{p}_A$ is attached, occurs in the plane spanned by $\mathbf{p}_A$ and $\mathbf{E}$. Thus, the rotation rate density $\bm{\Omega}_A^{\,\text{m}}$ is collinear to the torque $\mathbf{p}_A\times\mathbf{E}$ and $\mathsf{L}_{AA}^{\text{m}\text{m}} = L_{AA}^{\text{m}\text{m}}\,\mathbb{1}$. Substituting the relation~\eqref{relaxation el pol 0} into the time evolution equation~\eqref{eq el pol} for $\mathbf{p}_A$ yields an equation accounting for the Debye relaxation of the electric dipoles $\mathbf{p}_A$ in the presence of an electric field $\mathbf{E}$, i.e.
\begin{equation}\label{relaxation el pol}
\bm{\dot{\mathbf{p}}}_A = -\,\alpha_{A}\,\mathbf{p}_A\times\left(\mathbf{p}_A\times\mathbf{E}\right)\ ,
\end{equation}
where 
\begin{equation*}
\alpha_{A} = n_A^{-1}\,L_{AA}^{\text{m}\text{m}}\ ,
\end{equation*}
is a phenomenological friction coefficient.

The electric field $\mathbf{E}$ is an effective field that is defined with respect to the local infinitesimal system. It consists of two contributions, i.e. 
\begin{equation}\label{E}
\mathbf{E} = \mathbf{E}^{\,\text{ext}} + \mathbf{E}^{\,\text{int}}\ .
\end{equation}
The first contribution $\mathbf{E}^{\,\text{ext}}$ is due to an external field applied on the whole system. The second contribution $\mathbf{E}^{\,\text{int}}$ is due to the dipolar interaction with the infinitesimal systems that are in the neighbourhood of the local system. Note that this contribution is internal to the whole system, but external to the local infinitesimal system. In the neighbourhood of the local system, the Debye energy density due to the dipolar interaction between the local systems is proportional to the magnitude of the spatial variation of the electric polarisation. Thus, the Debye energy density can be expressed as,
\begin{equation}\label{Debye energy density}
-\,\mathbf{P}\cdot\mathbf{E} = -\,\mathbf{P}\cdot\mathbf{E}^{\,\text{ext}} -\,D\,\mathbf{P}\,\bm{\nabla}^2\mathbf{P}\ ,
\end{equation}
where $D$ is a phenomenological parameter. Note that the spatial variation is expressed in terms of a Laplacian $\bm{\nabla}^2$ and not a gradient $\bm{\nabla}$, since otherwise the contributions of local neighbouring systems located on opposite sides of the local system would cancel out. The expressions~\eqref{E} and~\eqref{Debye energy density} imply that the electric field $\mathbf{E}^{\,\text{int}}$ is given by, 
\begin{equation}\label{E int}
\mathbf{E}^{\,\text{int}} = D\,\bm{\nabla}^2\mathbf{P}\ .
\end{equation}
For a single homogeneous sample made of a single chemical substance $A$, consisting of electric dipoles $\mathbf{p}_A$, the electric polarisation $\mathbf{P} = n_A\,\mathbf{p}_A$. Thus, the time evolution equation~\eqref{relaxation el pol} becomes,
\begin{align}\label{relaxation el pol bis}
&\bm{\dot{\mathbf{p}}}_A = -\,\alpha_{A}\,\mathbf{p}_A\times\left(\mathbf{p}_A\times\mathbf{E}^{\,\text{ext}}\right)\\
&\phantom{\bm{\dot{\mathbf{p}}}_A =} -\,D\,\alpha_{A}\,\mathbf{p}_A\times\Big(\mathbf{p}_A\times\bm{\nabla}^2\left(n_A\,\mathbf{p}_A\right)\Big)\nonumber\ ,
\end{align}
where the terms on the RHS describe the relaxation due to an external electric field $\mathbf{E}^{\,\text{ext}}$ and to the interaction with the electric dipoles in the neighbourhood of the local system.

\subsubsection{Relaxation of magnetic dipoles and magnetisation waves}

\label{Relaxation of magnetic dipoles}

The first model of relaxation of magnetic dipoles $\mathbf{m}_A$ in the presence of a magnetic induction field $\mathbf{B}$ is due to Landau and Lifshitz~\cite{Landau:1935}. Later, Gilbert~\cite{Gilbert:2004} developed a Lagrangian formulation of this damping. Here, we show how our formalism accounts for this relaxation and for the magnetisation waves, which are the classical counterpart of the ``spin waves''~\cite{Saitoh:2012}.

We consider a homogeneous sample made of a single chemical substance $A$, consisting of magnetic dipoles $\mathbf{m}_A$ in the absence of an electric field, i.e. $\mathbf{E} = \mathbf{0}$. In the frame of sample, the chemical substance $A$ has no translational motion, i.e. $\mathbf{j}_A=\mathbf{0}$.

The absence of an electric field, i.e. $\mathbf{E} = \mathbf{0}$, implies that the linear phenomenological relation~\eqref{Onsager pseudo-vectors bis} reduces to,
\begin{equation}\label{relaxation mag 0}
\bm{\Omega}_A^{\,\mathbf{M}} = \mathsf{L}_{AA}^{\mathbf{M}\mathbf{M}}\cdot\left(\mathbf{m}_A\times\mathbf{B}\right)\ . 
\end{equation}
The process is irreversible and dissipative, which means that the Larmor energy $-\,\mathbf{m}_A\cdot\mathbf{B}$ has to diminish. This implies that the intrinsic rotation occurs in the plane spanned by $\mathbf{m}_A$ and $\mathbf{B}$. Thus, the rotation rate density $\bm{\Omega}_A^{\,\mathbf{M}}$ is collinear to the torque $\mathbf{m}_A\times\mathbf{B}$ and $\mathsf{L}_{AA}^{\mathbf{M}\mathbf{M}} = L_{AA}^{\mathbf{M}\mathbf{M}}\,\mathbb{1}$. Substituting the relation~\eqref{relaxation mag 0} into the time evolution equation~\eqref{eq mag} for $\mathbf{m}_A$ yields the Landau-Lifshitz equation~\cite{Gilbert:2004} accounting for the relaxation of the magnetic dipoles $\mathbf{m}_A$ in the presence of a magnetic induction field $\mathbf{B}$, i.e.
\begin{equation}\label{relaxation mag}
\bm{\dot{\mathbf{m}}}_A = \gamma_{A}\,\mathbf{m}_A\times\mathbf{B}-\,\beta_{A}\,\mathbf{m}_A\times\left(\mathbf{m}_A\times\mathbf{B}\right)\ ,
\end{equation}
where 
\begin{equation*}
\beta_{A} = n_A^{-1}\,L_{AA}^{\mathbf{M}\mathbf{M}}\ ,
\end{equation*}
is a phenomenological friction coefficient. Note that the relaxation of the magnetic dipoles is due to a mix of bodily rotation~\cite{Martsenyuk:1974} composed with Néel-type magnetic relaxation~\cite{Raikher:2004}.

The magnetic induction field $\mathbf{B}$ is an effective field that is defined with respect to the local infinitesimal system. It consists of two contributions, i.e. 
\begin{equation}\label{B}
\mathbf{B} = \mathbf{B}^{\,\text{ext}} + \mathbf{B}^{\,\text{int}}\ .
\end{equation}
The first contribution $\mathbf{B}^{\,\text{ext}}$ is due to an external field applied on the whole system. The second contribution $\mathbf{B}^{\,\text{int}}$ is due to the ferromagnetic interaction with the infinitesimal systems that are in the neighbourhood of the local system. In the neighbourhood of the local system, the Larmor energy density due to the ferromagnetic interaction between the local systems is proportional to the magnitude of the spatial variation of the magnetisation~\cite{Kittel:1949}. Thus, the Larmor energy density can be expressed as,
\begin{equation}\label{Larmor energy density}
-\,\mathbf{M}\cdot\mathbf{B} = -\,\mathbf{M}\cdot\mathbf{B}^{\,\text{ext}} -\,A\,\mathbf{M}\,\bm{\nabla}^2\mathbf{M}\ ,
\end{equation}
where $A$ is a phenomenological parameter called the stiffness constant~\cite{Saitoh:2012}. The expressions~\eqref{B} and~\eqref{Larmor energy density} imply that the magnetic induction field $\mathbf{B}^{\,\text{int}}$ is given by, 
\begin{equation}\label{B int}
\mathbf{B}^{\,\text{int}} = A\,\bm{\nabla}^2\mathbf{M}\ .
\end{equation}
For a single homogeneous sample made of a single chemical substance $A$, consisting of magnetic dipoles $\mathbf{m}_A$, the magnetisation $\mathbf{M} = n_A\,\mathbf{m}_A$. Thus, the Landau-Lifshitz equation~\eqref{relaxation mag} becomes,
\begin{align}\label{relaxation mag bis}
&\bm{\dot{\mathbf{m}}}_A = \gamma_{A}\,\mathbf{m}_A\times\mathbf{B}^{\,\text{ext}}-\,\beta_{A}\,\mathbf{m}_A\times\left(\mathbf{m}_A\times\mathbf{B}^{\,\text{ext}}\right)\nonumber\\
&\phantom{\bm{\dot{\mathbf{m}}}_A =} + A\,\gamma_{A}\,\mathbf{m}_A\times\bm{\nabla}^2\left(n_A\,\mathbf{m}_A\right)\\
&\phantom{\bm{\dot{\mathbf{m}}}_A =} -\,A\,\beta_{A}\,\mathbf{m}_A\times\Big(\mathbf{m}_A\times\bm{\nabla}^2\left(n_A\,\mathbf{m}_A\right)\Big)\nonumber\ .
\end{align}
The first and second terms on the RHS of the Landau-Lifschitz equation~\eqref{relaxation mag bis} describe respectively the precession and relaxation due to an external magnetic induction field $\mathbf{B}^{\,\text{ext}}$. The third term describes magnetisation waves~\cite{Kittel:1951} and the fourth term describes the relaxation of magnetisation waves.

\subsubsection{Thermally driven magnetisation current}

\label{Thermally driven magnetisation current}

The relaxation of magnetic dipoles in a ferroelectric metal, established in Sec.~\ref{Relaxation of magnetic dipoles}, rests on the assumption that the chemical units carrying the magnetic dipoles are at rest. Here, we consider a ferromagnetic conductor made of a fixed lattice and conduction electrons. The magnetic dipoles are carried by the core electrons of the lattice and the conduction electrons (e.g. `d-f' and `s-p' electrons respectively). On the scale of interest, the core electrons of the lattice are described as a continuum of chemical type $A$ at rest and the conduction electrons as a fluid of chemical type $B$. The core electrons carry a magnetic dipole $\mathbf{m}_A$ and the conduction electrons a magnetic dipole $\mathbf{m}_B$. A quasi-uniform and constant magnetic induction field $\mathbf{B}$ and temperature gradient $\bm{\nabla}\,T$ are applied on the ferromagnetic conductor. No electric field is applied, i.e. $\mathbf{E} = -\,\bm{\nabla}\,V = \mathbf{0}$.

Since the core electrons are at rest on the scale of interest, the current density of the core electron continuum vanishes, i.e. $\mathbf{j}_A=\mathbf{0}$. In our formalism, we neglect the ``chemical reactions'' $a$ between the core and conduction electrons, i.e. $\omega_a=0$. Thus, the time evolution equation~\eqref{eq mag} for the magnetic dipoles $\mathbf{m}_A$ of the core electron continuum $A$ reduces to,
\begin{equation}\label{eq mag el core 0}
\begin{split}	
&n_A\,\bm{\dot{\mathbf{m}}}_A = \gamma_A\,n_A\left(\mathbf{m}_A\times\mathbf{B}\right) -\,\mathbf{m}_A\times\bm{\Omega}_A^{\,\mathbf{M}}\\
&\phantom{n_A\,\bm{\dot{\mathbf{m}}}_A = } + \gamma_{AB}\,n_A\left(\mathbf{m}_A\times n_B\,\mathbf{m}_B\right)\ .
\end{split}
\end{equation}
The absence of electric field, i.e. $\mathbf{E} = \mathbf{0}$, implies that the linear phenomenological relation~\eqref{Onsager pseudo-vectors} for the core electron continuum reduces to,
\begin{equation}\label{irr mag 1}
\bm{\Omega}_A^{\,\mathbf{M}} = \mathsf{L}_{AA}^{\mathbf{M}\mathbf{M}}\cdot\left(\mathbf{m}_A\times\mathbf{B}\right) + \mathsf{L}_{AB}^{\mathbf{M}\mathbf{M}}\cdot\left(\mathbf{m}_B\times\mathbf{B}\right)\ . 
\end{equation}
The effect on the rotation of the core electrons of the magnetic torque $\mathbf{m}_B\times\mathbf{B}$ of the conduction electrons is negligible compared to the effect of the magnetic torque $\mathbf{m}_A\times\mathbf{B}$ of the core electrons, i.e. $\mathsf{L}_{AB}\ll\mathsf{L}_{AA}$, which yields the phenomenological equation~\eqref{relaxation mag 0}. The magnetic induction field $\mathbf{B}$ consists of two contributions as established in equation~\eqref{B} where the magnetisation $\mathbf{M} = n_A\,\mathbf{m}_A + n_B\,\mathbf{m}_B$, i.e. 
\begin{equation}\label{B bis}
\mathbf{B} = \mathbf{B}^{\,\text{ext}} + A\,\bm{\nabla}^2\left(n_A\,\mathbf{m}_A + n_B\,\mathbf{m}_B\right)\ .
\end{equation}
Following the same lines of thought as in Sec.~\ref{Relaxation of magnetic dipoles}, the local time evolution equation~\eqref{eq mag el core 0} for the magnetic dipoles of the core electron continuum is found to be,
\begin{align}\label{eq mag el core}
&\bm{\dot{\mathbf{m}}}_A = \gamma_A\left(\mathbf{m}_A\times\mathbf{B}^{\,\text{ext}}\right) -\,\beta_{A}\,\mathbf{m}_A\times\left(\mathbf{m}_A\times\mathbf{B}^{\,\text{ext}}\right)\nonumber\\
&\phantom{\bm{\dot{\mathbf{m}}}_A =} + A\,\gamma_{A}\,\mathbf{m}_A\times\bm{\nabla}^2\left(n_A\,\mathbf{m}_A + n_B\,\mathbf{m}_B\right)\nonumber\\
&\phantom{\bm{\dot{\mathbf{m}}}_A =} -\,A\,\beta_{A}\,\mathbf{m}_A\times\Big(\mathbf{m}_A\times\bm{\nabla}^2\left(n_A\,\mathbf{m}_A + n_B\,\mathbf{m}_B\right)\Big)\nonumber\\
&\phantom{\bm{\dot{\mathbf{m}}}_A = } + \gamma_{AB}\left(\mathbf{m}_A\times n_B\,\mathbf{m}_B\right)\ .
\end{align}
The last term on the RHS of the equation~\eqref{eq mag el core} is expected to lead to the magnetisation transfer torque, which is the classical counterpart of the ``spin transfer torque'', generated by the magnetic dipoles $\mathbf{m}_B$ of the conduction electron fluid on the magnetic dipoles $\mathbf{m}_A$ of the core electron continuum~\cite{Ansermet:2010}. Note that the time evolution of the magnetic dipoles $\mathbf{m}_A$ of the core electron continuum has no explicit dependence on the temperature gradient because these electrons do not undergo transport. However, as shown below, the dynamics of the magnetic dipoles $\mathbf{m}_B$ depends on the temperature gradient.

Since we neglect the ``chemical reactions'' $a$ between the core and conduction electrons, i.e. $\omega_a=0$, the time evolution equation~\eqref{eq mag} for the magnetic dipoles $\mathbf{m}_B$ of the core electron continuum $B$ is given by,
\begin{align}\label{eq mag el cconduction 0}
&n_B\,\bm{\dot{\mathbf{m}}}_B = \gamma_B\,n_B\left(\mathbf{m}_B\times\mathbf{B}\right) -\,\mathbf{m}_B\times\bm{\Omega}_B^{\,\mathbf{M}} -\,\left(\mathbf{j}_B\!\cdot\!\bm{\nabla}\right)\mathbf{m}_B\nonumber\\
&\phantom{n_B\,\bm{\dot{\mathbf{m}}}_B = } + \gamma_{BA}\,n_B\left(\mathbf{m}_B\times n_A\,\mathbf{m}_A\right)\ .
\end{align}
Following the same procedure as for the core electrons, we recast the local time evolution equation~\eqref{eq mag el core 0} for the magnetic dipoles of the conduction electron fluid as,
\begin{align}\label{eq mag el conduction}
&\bm{\dot{\mathbf{m}}}_B = \gamma_B\left(\mathbf{m}_B\times\mathbf{B}^{\,\text{ext}}\right) -\,\beta_{B}\,\mathbf{m}_B\times\left(\mathbf{m}_B\times\mathbf{B}^{\,\text{ext}}\right)\nonumber\\
&\phantom{\bm{\dot{\mathbf{m}}}_B =} + A\,\gamma_{B}\,\mathbf{m}_B\times\bm{\nabla}^2\left(n_A\,\mathbf{m}_A + n_B\,\mathbf{m}_B\right)\\
&\phantom{\bm{\dot{\mathbf{m}}}_B =} -\,A\,\beta_{B}\,\mathbf{m}_B\times\Big(\mathbf{m}_B\times\bm{\nabla}^2\left(n_A\,\mathbf{m}_A + n_B\,\mathbf{m}_B\right)\Big)\nonumber\\
&\phantom{\bm{\dot{\mathbf{m}}}_B = }  -\,n_B^{-1}\left(\mathbf{j}_B\cdot\bm{\nabla}\right)\mathbf{m}_B + \gamma_{BA}\left(\mathbf{m}_B\times n_A\,\mathbf{m}_A\right)\ .\nonumber
\end{align}
In order to find an explicit expression for the third term on the RHS of the time evolution equation~\eqref{eq mag el conduction}, we describe the transport of the conduction electron fluid in the presence of a uniform magnetic induction field $\mathbf{B}$ and a temperature gradient $\bm{\nabla}\,T$. In the absence of an electric field, i.e. $\bm{\nabla}\,V = \mathbf{0}$, the linear phenomenological relation~\eqref{Onsager vectors bis} reduces to,
\begin{equation}\label{lin phen}
\mathbf{j}_B = -\,\frac{1}{q_B}\,\bm{\sigma}_B\cdot\bm{\varepsilon}_B\cdot\bm{\nabla}\,T\ ,
\end{equation}
where the isothermal electric conductivity tensor $\bm{\sigma}_B$ and the thermopower tensor $\bm{\varepsilon}_B$ are defined respectively as~\cite{Callen:1960},
\begin{equation}\label{sigma epsilon}
\begin{split}
&\bm{\sigma}_B \equiv q_B^2\,\mathsf{L}_{BB}\ ,\\
&\bm{\varepsilon}_B \equiv \frac{1}{q_B}\,\mathsf{L}_{BB}^{-1}\cdot\mathsf{L}_{Bs}\ ,
\end{split}
\end{equation}
and satisfy the following symmetries~\cite{Landau:1982},
\begin{equation*}
\begin{split}
&\bm{\sigma}_B\left(\mathbf{B}\right) = \bm{\sigma}_B^{T}\left(-\,\mathbf{B}\right)\ ,\\
&\bm{\varepsilon}_B\left(\mathbf{B}\right) = \bm{\varepsilon}_B^{T}\left(-\,\mathbf{B}\right)\ ,
\end{split}
\end{equation*}
where the exponent $^{T}$ stands for transpose. Thus, these phenomenological tensors can be split into symmetric and antisymmetric parts according to,
\begin{equation}\label{sigma epsilon split}
\begin{split}
&\bm{\sigma}_B\cdot\mathbf{a} = \sigma_{B\parallel}\,\mathbf{a} + \sigma_{B\perp}\left(\mathbf{\hat{B}}\times\mathbf{a}\right)\ ,\\
&\bm{\varepsilon}_B\cdot\mathbf{a} = \varepsilon_{B\parallel}\,\mathbf{a} + \varepsilon_{B\perp}\left(\mathbf{\hat{B}}\times\mathbf{a}\right)\ ,
\end{split}
\end{equation}
where the magnetic induction field $\mathbf{B} = B\,\mathbf{\hat{B}}$ and $\mathbf{a}$ is an arbitrary vector. By symmetry the collinear factors $\sigma_{B\parallel}$ and $\varepsilon_{B\parallel}$ are independent of the intensity $B$ of the magnetic induction field. The Hall and Nernst effects~\cite{Landau:1982,Callen:1960} require the orthogonal factors $\sigma_{B\perp}$ and $\varepsilon_{B\perp}$ respectively to be inversely proportional to $B$, i.e. $\sigma_{B\perp} \propto B^{-1}$ and $\varepsilon_{B\perp} \propto B^{-1}$. Using the splittings~\eqref{sigma epsilon split}, the linear phenomenological relation~\eqref{lin phen} becomes,
\begin{equation}\label{lin phen rel}
\mathbf{j}_B = -\,\frac{\sigma_{B\parallel}\,\varepsilon_{B\parallel}}{q_B}\,\bm{\nabla}\,T -\,\frac{\sigma_{B\parallel}\,\varepsilon_{B\perp} + \sigma_{B\perp}\,\varepsilon_{B\parallel}}{q_B}\left(\mathbf{\hat{B}}\times\bm{\nabla}\,T\right)
\end{equation}
where the first term on the RHS describes the Soret effect~\cite{Callen:1960} for the component of the magnetic induction field $\mathbf{B}$ that is collinear to the temperature gradient $\bm{\nabla}\,T$ and does not affect the transport, and the second term describes the Ettingshausen effect~\cite{Callen:1960} per unit of electric charge $q_B$ for the components of the magnetic induction field $\mathbf{B}$ that are orthogonal to the temperature gradient $\bm{\nabla}\,T$ and affect the transport.

Since the core electrons $A$ do not participate to the transport, the magnetisation current density tensor $\mathsf{j}_{\mathbf{M}}$ is entirely due to the transport of the conduction electrons $B$. Using the relations~\eqref{mag cur dens} and~\eqref{lin phen rel}, $\mathsf{j}_{\mathbf{M}}$ is found to be,
\begin{equation}\label{therm mag curr}
\begin{split}
&\mathsf{j}_{\mathbf{M}} = -\,\frac{\sigma_{B\parallel}\,\varepsilon_{B\parallel}}{q_B}\,\mathbf{m}_B\odot\bm{\nabla}\,T \\
&\phantom{\mathsf{j}_{\mathbf{M}} = } -\,\frac{\sigma_{B\parallel}\,\varepsilon_{B\perp} + \sigma_{B\perp}\,\varepsilon_{B\parallel}}{q_B}\,\mathbf{m}_B\odot\left(\mathbf{\hat{B}}\times\bm{\nabla}\,T\right)\ .
\end{split}
\end{equation}

Substituting the linear phenomenological equation~\eqref{lin phen rel} into the local time evolution equation~\eqref{eq mag el conduction} for the magnetic dipoles $\mathbf{m}_B$ of the conduction electron fluid $B$, the latter becomes,
\begin{align}\label{eq mag el conduction para}
&\bm{\dot{\mathbf{m}}}_B = \gamma_B\left(\mathbf{m}_B\times\mathbf{B}^{\,\text{ext}}\right) -\,\beta_{B}\,\mathbf{m}_B\times\left(\mathbf{m}_B\times\mathbf{B}^{\,\text{ext}}\right) \vphantom{\frac{\sigma_{B\parallel}\,\varepsilon_{B\parallel}}{q_B\,n_B}}\nonumber\\
&\phantom{\bm{\dot{\mathbf{m}}}_B =} + A\,\gamma_{B}\,\mathbf{m}_B\times\bm{\nabla}^2\left(n_A\,\mathbf{m}_A + n_B\,\mathbf{m}_B\right)\\
&\phantom{\bm{\dot{\mathbf{m}}}_B =} -\,A\,\beta_{B}\,\mathbf{m}_B\times\Big(\mathbf{m}_B\times\bm{\nabla}^2\left(n_A\,\mathbf{m}_A + n_B\,\mathbf{m}_B\right)\Big)\nonumber\\
&\phantom{\bm{\dot{\mathbf{m}}}_B = }  + \gamma_{BA}\left(\mathbf{m}_B\times n_A\,\mathbf{m}_A\right) + \frac{\sigma_{B\parallel}\,\varepsilon_{B\parallel}}{q_B\,n_B}\,\bm{\nabla}\,T\cdot\bm{\nabla}\,\mathbf{m}_B\nonumber\\
&\phantom{\bm{\dot{\mathbf{m}}}_B = }  + \frac{\sigma_{B\parallel}\,\varepsilon_{B\perp} + \sigma_{B\perp}\,\varepsilon_{B\parallel}}{q_B\,n_B}\,\left(\mathbf{\hat{B}}\times\bm{\nabla}\,T\right)\cdot\bm{\nabla}\,\mathbf{m}_B\ ,\nonumber
\end{align}
where the first and second terms on the RHS account respectively for the precession and the relaxation of the magnetic dipoles $\mathbf{m}_B$. The third and fourth terms account respectively for the magnetisation waves and the relaxation of the magnetisation waves. The fifth term accounts accounts for the interaction between the magnetic dipoles $\mathbf{m}_A$ and $\mathbf{m}_B$ of the core and conduction electrons respectively. The sixth and seventh terms account for the transport of $\mathbf{m}_B$ and describe the magnetisation accumulation, which is the classical counterpart of the ``spin accumulation'', generated by the temperature gradient $\bm{\nabla}\,T$ in the presence of a magnetic induction field $\mathbf{B}$. The sixth term describes the magnetisation accumulation collinear to the temperature gradient and the seventh term describes the magnetisation accumulation orthogonal to the temperature gradient and the magnetic induction field.

The times scales associated to the precession and the relaxation of the conduction electrons are much smaller than the time scales associated to the magnetisation accumulation and the magnetisation transfer torque. Thus, on the latter time scales, the first, second, third and fourth terms on the RHS of the time evolution equation~\eqref{eq mag el conduction para} can be neglected, i.e.
\begin{equation}\label{eq mag el conduction para bis}
\begin{split}
&\bm{\dot{\mathbf{m}}}_B = 	\gamma_{BA}\left(\mathbf{m}_B\times n_A\,\mathbf{m}_A\right) + \frac{\sigma_{B\parallel}\,\varepsilon_{B\parallel}}{q_B\,n_B}\,\bm{\nabla}\,T\cdot\bm{\nabla}\,\mathbf{m}_B\\
&\phantom{\bm{\dot{\mathbf{m}}}_B = }  + \frac{\sigma_{B\parallel}\,\varepsilon_{B\perp} + \sigma_{B\perp}\,\varepsilon_{B\parallel}}{q_B\,n_B}\,\left(\mathbf{\hat{B}}\times\bm{\nabla}\,T\right)\cdot\bm{\nabla}\,\mathbf{m}_B\ .  
\end{split}
\end{equation}

\subsubsection{Electrically driven magnetisation current}

The time evolution of the magnetic dipoles of a ferromagnetic conductor, made of a fixed lattice and conduction electrons, in the presence of a quasi-uniform magnetic field $\mathbf{B}$ and a temperature gradient $\bm{\nabla}\,T$ was established in Sec.~\ref{Thermally driven magnetisation current}. Here we consider an isothermal conductor, i.e. $\bm{\nabla}\,T = \mathbf{0}$, in the presence of a quasi-uniform electric field, i.e. $\mathbf{E} = -\,\bm{\nabla}\,V$. On the scale of interest, the core electrons of the lattice are described as a continuum of chemical type $A$ at rest and the conduction electrons as a fluid of chemical type $B$. The core electrons carry a magnetic dipole $\mathbf{m}_A$ and the conduction electrons a magnetic dipole $\mathbf{m}_B$.

The core electron continuum is at rest and the local time evolution equation~\eqref{eq mag el core} for the magnetic dipoles $\mathbf{m}_A$ is independent of the transport. The time evolution for the magnetic dipoles $\mathbf{m}_B$ of the conduction electron fluid is given by the equation~\eqref{eq mag el conduction}. 

In order to find an explicit expression for the third term on the RHS of the time evolution equation~\eqref{eq mag el conduction}, we describe the transport of the conduction electron fluid in the presence of a magnetic induction field $\mathbf{B}$ and a electric potential gradient $\bm{\nabla}\,V$. In the absence of a temperature gradient, i.e. $\bm{\nabla}\,T = \mathbf{0}$, the linear phenomenological relation~\eqref{Onsager vectors bis} reduces to,
\begin{equation}\label{lin phen el}
\mathbf{j}_B = -\,\frac{1}{q_B}\,\bm{\sigma}_B\cdot\bm{\nabla}\,V\ .
\end{equation}
Using the first splitting~\eqref{sigma epsilon split}, the linear phenomenological relation~\eqref{lin phen el} becomes,
\begin{equation}\label{lin phen rel el}
\mathbf{j}_B = -\,\frac{\sigma_{B\parallel}}{q_B}\,\bm{\nabla}\,V -\,\frac{\sigma_{B\perp}}{q_B}\left(\mathbf{\hat{B}}\times\bm{\nabla}\,V\right)
\end{equation}
where the first term on the RHS is Ohm's law~\cite{Callen:1960} per unit of electric charge $q_B$ for the component of the magnetic induction field $\mathbf{B}$ that is collinear to the electric potential gradient $\bm{\nabla}\,V$ and does not affect the transport, and the second term describes the Hall effect~\cite{Callen:1960} per unit of electric charge $q_B$ for the components of the magnetic induction field $\mathbf{B}$ that are orthogonal to the electric potential gradient $\bm{\nabla}\,V$ and affect the transport.

Since the core electrons $A$ do not participate to the transport, the magnetisation current density tensor $\mathsf{j}_{\mathbf{M}}$ is entirely due to the transport of the conduction electrons $B$. Using the relations~\eqref{mag cur dens} and~\eqref{lin phen rel el}, $\mathsf{j}_{\mathbf{M}}$ is found to be,
\begin{equation}\label{el mag curr}
\mathsf{j}_{\mathbf{M}} = -\,\frac{\sigma_{B\parallel}}{q_B}\,\mathbf{m}_B\odot\bm{\nabla}\,V -\,\frac{\sigma_{B\perp}}{q_B}\,\mathbf{m}_B\odot\left(\mathbf{\hat{B}}\times\bm{\nabla}\,V\right)\ .
\end{equation}

Substituting the linear phenomenological equation~\eqref{lin phen rel el} into the local time evolution equation~\eqref{eq mag el conduction} for the magnetic dipoles $\mathbf{m}_B$ of the conduction electron fluid $B$, the latter becomes,
\begin{align}\label{eq mag el conduction para el}
&\bm{\dot{\mathbf{m}}}_B = \gamma_B\left(\mathbf{m}_B\times\mathbf{B}^{\,\text{ext}}\right) -\,\beta_{B}\,\mathbf{m}_B\times\left(\mathbf{m}_B\times\mathbf{B}^{\,\text{ext}}\right)\nonumber\\
&\phantom{\bm{\dot{\mathbf{m}}}_B =} + A\,\gamma_{B}\,\mathbf{m}_B\times\bm{\nabla}^2\left(n_A\,\mathbf{m}_A + n_B\,\mathbf{m}_B\right)\\
&\phantom{\bm{\dot{\mathbf{m}}}_B =} -\,A\,\beta_{B}\,\mathbf{m}_B\times\Big(\mathbf{m}_B\times\bm{\nabla}^2\left(n_A\,\mathbf{m}_A + n_B\,\mathbf{m}_B\right)\Big)\nonumber\\
&\phantom{\bm{\dot{\mathbf{m}}}_B = } + \gamma_{BA}\left(\mathbf{m}_B\times n_A\,\mathbf{m}_A\right) + \frac{\sigma_{B\parallel}}{q_B\,n_B}\,\bm{\nabla}\,V\cdot\bm{\nabla}\,\mathbf{m}_B\nonumber\\
&\phantom{\bm{\dot{\mathbf{m}}}_B = } + \frac{\sigma_{B\perp}}{q_B\,n_B}\left(\mathbf{\hat{B}}\times\bm{\nabla}\,V\right)\cdot\bm{\nabla}\,\mathbf{m}_B\ ,\nonumber
\end{align}
where the sixth and seventh terms account for the transport of $\mathbf{m}_B$ and describe the magnetisation accumulation generated by the electric potential gradient $\bm{\nabla}\,V$ in the presence of a magnetic induction field $\mathbf{B}$. The seventh term describes the magnetisation accumulation collinear to the electric potential gradient and the fifth term describes the magnetisation accumulation orthogonal to the electric potential gradient and the magnetic induction field.

On the time scales associated to the magnetisation accumulation and the magnetisation transfer torque, the time evolution equation~\eqref{eq mag el conduction para el} reduces to, i.e.
\begin{equation}\label{eq mag el conduction para bis el}
\begin{split}
&\bm{\dot{\mathbf{m}}}_B = 	\gamma_{BA}\left(\mathbf{m}_B\times n_A\,\mathbf{m}_A\right) + \frac{\sigma_{B\parallel}}{q_B\,n_B}\,\bm{\nabla}\,V\cdot\bm{\nabla}\,\mathbf{m}_B\\
&\phantom{\bm{\dot{\mathbf{m}}}_B = } + \frac{\sigma_{B\perp}}{q_B\,n_B}\left(\mathbf{\hat{B}}\times\bm{\nabla}\,V\right)\cdot\bm{\nabla}\,\mathbf{m}_B\ .  
\end{split}
\end{equation}

\subsubsection{Thermodynamics of magnetic vortices}

The magnetic counterpart of electrically polarised liquid crystals are magnetic vortices or skyrmions~\cite{Skyrme:1962}. Skyrmions were observed recently~\cite{Seki:2012} in the insulating ferromagnet Cu$_2$OSeO$_3$. Our formalism predicts the precession and relaxation of magnetic vortices in the presence of a temperature gradient.

We consider magnetic vortices that can be treated as a continuum. They form a gyromagnetic medium with a vorticity axis of unit vector $\mathbf{\hat{n}}$. We assume that $\mathbf{\hat{n}}$ is uniform. The magnetic vortices are made of ions and core electrons and are considered as an electrically neutral substance of type $A$, i.e. $q_A = 0$. In the absence of an external interaction, the vortices have no net magnetisation. On a macroscopic scale, the continuum is homogeneous, i.e. $\bm{\nabla}\,\mu_A = \mathbf{0}$. We assume that the viscosity can be neglected, i.e.  i.e. $\bm{\nabla}\,\mathbf{v}_A = \mathbf{0}$ and $\bm{\nabla}\,\bm{\omega}_A = \mathbf{0}$. Moreover, since the magnetic dipoles are carried by the electrons, that have no intrinsic angular mass, we do not consider the rotation of the matter but we take into account only the rotation of the magnetisation. In an electrical insulator, the magnetic vortices have no translational motion, i.e. $\mathbf{j}_A = \mathbf{0}$. Furthermore, a temperature gradient $\bm{\nabla}\,T$ is applied along a direction that is different from the vorticity axis of the magnetic vortices. 

In this case, the linear phenomenological relation~\eqref{Onsager vectors bis} reduces to,
\begin{equation}\label{Mag Seebeck effect 00}
\mathbf{m}_A\,\bm{\nabla}\,\mathbf{B} = \left(\mathsf{L}_{AA}^{-1}\cdot\mathsf{L}_{As}\right)\cdot\bm{\nabla}\,T\ ,
\end{equation}
where the temperature gradient $\bm{\nabla}\,T$ induces a magnetic induction field $\mathbf{B}$ and the magnetic dipoles $\mathbf{m}_A$ rotate in an asymmetric manner in order to lower the Larmor energy $-\,\mathbf{m}_A\cdot\mathbf{B}$. Thus, a net magnetisation $\mathbf{M}$ is generated along the vorticity axis $\mathbf{\hat{n}}$, i.e.
\begin{equation}\label{mag axis}
\mathbf{M} = n_A\,\mathbf{m}_A = n_A\left(\mathbf{m}_A\cdot\mathbf{\hat{n}}\right)\mathbf{\hat{n}}\ ,
\end{equation}
and the linear relation~\eqref{Mag Seebeck effect 00} is recast as,
\begin{equation}\label{Mag Seebeck effect 0}
\mathbf{M}\,\bm{\nabla}\,\mathbf{B} = \left(n_A\,\mathsf{L}_{AA}^{-1}\cdot\mathsf{L}_{As}\right)\cdot\bm{\nabla}\,T\ .
\end{equation}
The spatial symmetry requires the magnetisation force $\mathbf{M}\,\bm{\nabla}\,\mathbf{B}$ and the thermal force $-\,\bm{\nabla}\,T$ to be collinear, i.e.
\begin{equation}\label{Mag Seebeck effect 0 bis}
\mathbf{M}\,\bm{\nabla}\,\mathbf{B} = n_A\,L_{AA}^{-1}\,L_{As}\,\bm{\nabla}\,T\ ,
\end{equation}
where $\mathsf{L}_{AA}^{-1} = L_{AA}^{-1}\,\mathbb{1}$ and $\mathsf{L}_{As} = L_{As}\,\mathbb{1}$. Thus, the phenomenological relation~\eqref{Mag Seebeck effect 0 bis} can be recast as,
\begin{equation}\label{Mag Seebeck effect 0 ter}
\mathbf{M}\,\bm{\nabla}\,\mathbf{B} = \lambda\,n_A\,k_B\,\bm{\nabla}\,T\ ,
\end{equation}
where $\lambda$ is a dimensionless parameter by the relation~\eqref{en equipartition}. The magnetisation force density $\mathbf{M}\,\bm{\nabla}\,\mathbf{B}$ satisfies the vectorial identity,
\begin{align}\label{rel vect Mag Lehmann}
&\mathbf{M}\,\bm{\nabla}\,\mathbf{B} =  \bm{\nabla}\,\left(\mathbf{M}\cdot\mathbf{B}\right)  -\,\mathbf{B}\,\bm{\nabla}\,\mathbf{M} -\,\mathbf{B}\times\left(\bm{\nabla}\times\mathbf{M}\right)\\
& =  -\,\bm{\nabla}_{\mathbf{M}}\cdot\left(\mathbf{M}\odot\mathbf{B} -\,\left(\mathbf{M}\cdot\mathbf{B}\right)\mathbb{1}\right) + \left(\bm{\nabla}\times\mathbf{M}\right)\times\mathbf{B}\nonumber
\end{align}
where we used Thomson's law, i.e.
\begin{equation}\label{Thomson law}
\bm{\nabla}\cdot\mathbf{B} = 0\ ,
\end{equation}
and the index $\vphantom{a}_\mathbf{M}$ denotes that there is a dot product between the covariant differential operator $\bm{\nabla}$ and the magnetisation $\mathbf{M}$. The term $\bm{\nabla}_{\mathbf{M}}\cdot\left(\mathbf{M}\odot\mathbf{B} -\,\left(\mathbf{M}\cdot\mathbf{B}\right)\mathbb{1}\right)$ in the identity~\eqref{rel vect Mag Lehmann} corresponds to a surface contribution after integration over the volume of the continuum of magnetic vortices~\cite{Reuse:2012}. Thus, it can be neglected in the bulk of the continuum where the magnetisation force density $\mathbf{M}\,\bm{\nabla}\,\mathbf{B}$ is expressed as,
\begin{equation}\label{mag force dens}
\mathbf{M}\,\bm{\nabla}\,\mathbf{B} = \left(\bm{\nabla}\times\mathbf{M}\right)\times\mathbf{B}\ .
\end{equation}
Moreover, using the definition of the bound electric current density $\mathbf{j}_{\mathbf{M}}$, i.e.
\begin{equation}\label{Mag bound current density}
\mathbf{j}_{\mathbf{M}} \equiv \bm{\nabla}\times\mathbf{M}\ ,
\end{equation}
the magnetisation force density~\eqref{mag force dens} is recast as~\cite{Reuse:2012},
\begin{equation}\label{Mag rel vect Lehmann bis}
\mathbf{M}\,\bm{\nabla}\,\mathbf{B} = \mathbf{j}_{\mathbf{M}}\times\mathbf{B}\ ,
\end{equation}
which shows that it is the analog of the magnetic part of the Lorentz force density for bound currents. Substituting the expression~\eqref{Mag rel vect Lehmann bis} in the linear relation~\eqref{Mag Seebeck effect 0 bis}, the latter becomes, 
\begin{equation}\label{Thermomagnetic effect 0}
\mathbf{j}_{\mathbf{M}}\times\mathbf{B} = \lambda\,n_A\,k_B\,\bm{\nabla}\,T\ ,
\end{equation}
which implies that
\begin{equation}\label{Thermomagnetic effect 0 bis}
\mathbf{j}_{\mathbf{M}}\times\left(\mathbf{j}_{\mathbf{M}}\times\mathbf{B}\right) = \lambda\,n_A\,k_B\,\left(\mathbf{j}_{\mathbf{M}}\times\bm{\nabla}\,T\right)\ ,
\end{equation}
and that the thermally induced magnetic induction field $\mathbf{B}$ is orthogonal to the bound current $\mathbf{j}_{\mathbf{M}}$, i.e.
\begin{equation*}
\mathbf{j}_{\mathbf{M}}\cdot\mathbf{B} = 0\ .
\end{equation*}
Using the vectorial identity
\begin{equation*}
\mathbf{j}_{\mathbf{M}}\times\left(\mathbf{j}_{\mathbf{M}}\times\mathbf{B}\right) = \left(\mathbf{j}_{\mathbf{M}}\cdot\mathbf{B}\right)\mathbf{j}_{\mathbf{M}} -\,\mathbf{j}_{\mathbf{M}}^2\,\mathbf{B} = -\,\mathbf{j}_{\mathbf{M}}^2\,\mathbf{B}\ ,
\end{equation*}
the linear relation~\eqref{Thermomagnetic effect 0 bis} describing the magnetic field $\mathbf{B}$ induced by a temperature gradient $\bm{\nabla}\,T$ on magnetic vortices $\mathbf{j}_{\mathbf{M}}$ can be recast as a magnetic Seebeck effect for bound currents in an electric insulator, i.e.
\begin{equation}\label{Thermomagnetic effect ter}
\mathbf{B} = \bm{\varepsilon}_{\mathbf{M}}\times\bm{\nabla}\,T\ ,
\end{equation}
where the magnetic Seebeck vector for bound electric currents,
\begin{equation*}
\bm{\varepsilon}_{\mathbf{M}} \equiv -\,\lambda\,n_A\,k_B\,\mathbf{j}_{\mathbf{M}}^{-1}\ .
\end{equation*}
Note that the magnetic Seebeck effect~\eqref{Thermomagnetic effect ter} is the magnetic analog of the Seebeck effect~\eqref{Seebeck effect} in an electric insulator. 

Substituting the expression~\eqref{Thermomagnetic effect ter} for the induced magnetic field $\mathbf{B}$ into the local time evolution equation of the magnetic dipoles~\eqref{relaxation mag}, we obtain an expression for the thermally induced dynamics of the magnetic vortices, i.e.
\begin{align}\label{eq mag vortex}
&\bm{\dot{\mathbf{m}}}_A = \gamma_A\,\mathbf{m}_A\times\left(\bm{\varepsilon}_{\mathbf{M}}\times\bm{\nabla}\,T\right)\\
&\phantom{\bm{\dot{\mathbf{m}}}_A = }-\,\beta_A\,\mathbf{m}_A\times\Big(\mathbf{m}_A\times\left(\bm{\varepsilon}_{\mathbf{M}}\times\bm{\nabla}\,T\right)\Big)\ ,\nonumber
\end{align}
where the first term and second terms on the RHS describe respectively the precession and the relaxation of the magnetic dipoles of the core electrons of the magnetic vortices due to magnetic field $\mathbf{B}$ induced by a temperature gradient $\bm{\nabla}\,T$.

\subsubsection{Thermally driven magnetisation waves}

We consider a ferromagnetic insulator made of a fixed lattice with core electrons that carry a magnetic dipole $\mathbf{m}_A$. A constant external magnetic induction field $\mathbf{B}^{\,\text{ext}}$ and a temperature gradient $\bm{\nabla}\,T$ are applied. The time evolution of the magnetic dipoles is given by the Landau-Lifschitz equation~\eqref{relaxation mag}.

The magnetic induction field $\mathbf{B}$ appearing in the time evolution equation~\eqref{relaxation mag} is an effective field that is defined with respect to the local infinitesimal system. It consists of three contributions, i.e. 
\begin{equation}\label{B ter}
\mathbf{B} = \mathbf{B}^{\,\text{ext}} + \mathbf{B}^{\,\text{int}} + \mathbf{B}^{\,\text{ind}}\ .
\end{equation}
The first contribution $\mathbf{B}^{\,\text{ext}}$ is due to an external field applied on the whole system. The second contribution $\mathbf{B}^{\,\text{int}}$ is due to the ferromagnetic interaction with the infinitesimal systems that are in the neighbourhood of the local system. For a core electron continuum, the magnetisation $\mathbf{M} = n_A\,\mathbf{m}_A$ and the expression for $\mathbf{B}^{\,\text{int}}$ follows from equation~\eqref{B int}, i.e.
\begin{equation}\label{B int bis}
\mathbf{B}^{\,\text{int}} = A\,\bm{\nabla}^2\left(n_A\,\mathbf{m}_A\right)\ .
\end{equation}
The third contribution $\mathbf{B}^{\,\text{ind}}$ is induced by the temperature gradient $\bm{\nabla}\,T$. For a core electron continuum, the expression for $\mathbf{B}^{\,\text{ind}}$ follows from equation~\eqref{Thermomagnetic effect ter}, i.e.
\begin{equation}\label{Thermomagnetic effect quad}
\mathbf{B}^{\,\text{ind}} = \bm{\varepsilon}_{\mathbf{M}}\times\bm{\nabla}\,T\ ,
\end{equation}
where the bound current density yields,
\begin{equation}\label{b current}
\mathbf{j}_{\mathbf{M}} = \bm{\nabla}\times\left(n_A\,\mathbf{m}_A\right)\ .
\end{equation}
Note that the magnetic induction field $\mathbf{B}^{\,\text{ind}}$ is induced only in the presence of a bound current density $\mathbf{j}_{\mathbf{M}}$. This current density arises in the presence of magnetisation waves generated by the applied external magnetic field $\mathbf{B}^{\,\text{ext}}$.

Substituting the expressions~\eqref{B int bis} and~\eqref{Thermomagnetic effect quad} for the different contributions to the effective magnetic field $\mathbf{B}$ into the Landau-Lifschitz equation~\eqref{relaxation mag}, the latter becomes,
\begin{align}\label{therm magn eq}
&\bm{\dot{\mathbf{m}}}_A = \gamma_A\left(\mathbf{m}_A\times\mathbf{B}^{\,\text{ext}}\right) -\,\beta_{A}\,\mathbf{m}_A\times\left(\mathbf{m}_A\times\mathbf{B}^{\,\text{ext}}\right)\nonumber\\
&+ \gamma_{A}\,\mathbf{m}_A\times\Big(A\,\bm{\nabla}^2\left(n_A\,\mathbf{m}_A\right) + \bm{\varepsilon}_{\mathbf{M}}\times\bm{\nabla}\,T\Big)\\
&-\,\beta_{A}\,\mathbf{m}_A\!\times\!\bigg(\mathbf{m}_A\!\times\!\Big(A\,\bm{\nabla}^2\left(n_A\,\mathbf{m}_A\right) + \bm{\varepsilon}_{\mathbf{M}}\times\bm{\nabla}\,T\Big)\bigg)\ .\nonumber
\end{align}
where the third and fourth terms describe respectively how magnetisation waves and their relaxation are driven by a temperature gradient $\bm{\nabla}\,T$. The effect is maximal when the temperature gradient is orthogonal to the bound current $\mathbf{j}_{\mathbf{M}}$ generated by the applied external magnetic field $\mathbf{B}^{\,\text{ext}}$.


\section{Conclusion}

The thermodynamics of irreversible processes is considered for an electrically charged continuous medium containing spontaneous electric and magnetic dipoles in the presence of electromagnetic fields. Expressing the extensive matter state fields in terms of their chemical constituents yields explicit expressions for the current densities. Three types of irreversible terms are derived from these expressions. These are scalar, vectorial and pseudo-vectorial terms that describe respectively irreversible chemical processes, irreversible transport processes and irreversible relaxation processes. These processes are coupled through the time evolution equations. Note that with such an approach, the mathematical structure of the irreversible thermodynamics is uncovered physically without imposing it a priori using irreducible representations of the Euclidean group. 

As an illustration of our formalism, we describe notably the Lehmann and electric Lehmann effects, the relaxation of the electric and magnetic dipoles. In particular, we are able to predict the effect of a temperature gradient on the time evolution of the magnetic dipoles of conduction electrons interacting with core electrons, which leads to precise expressions for the thermal and electronic magnetisation accumulations. We also predict the precession and relaxation of magnetic vortices induced by a temperature gradient in the absence of an applied magnetic induction field, which is very innovative. Finally, in the presence of an applied magnetic induction field, we show explicitly how a temperature gradient drives magnetisation waves.


\begin{acknowledgement}

The authors would like to thank Fran\c{c}ois A. Reuse for theoretical guidance.

\end{acknowledgement}


\bibliography{references}
\bibliographystyle{epjc} 

\end{document}